\documentclass[pdflatex,sn-mathphys-num]{sn-jnl}

\pdfoutput=1
\usepackage{amsmath,amssymb}
\usepackage{graphicx}
\usepackage{booktabs}
\usepackage{makecell}
\usepackage{subcaption}

\begin{document}

\title{Cryogenic performance and long-term stability of Hamamatsu R8520-506 photomultiplier tubes in liquid argon}

\author[1,2]{Jilong Yin}\email{yinjl@ihep.ac.cn}

\author[1,3]{Gaoshuang Li}

\author[1,2]{Yike Shu}

\author*[1,2]{Yi Wang}\email{wangyi90@ihep.ac.cn}

\affil[1]{Institute of High Energy Physics, Chinese Academy of Sciences, \postcode{100049}, \city{Beijing}, \country{China}}

\affil[2]{University of Chinese Academy of Sciences, \postcode{10049}, \city{Beijing}, \country{China}}

\affil[3]{Zhengzhou University, \postcode{450001}, \city{Zhengzhou}, \country{China}}

\abstract{
\textbf{Purpose} Seven Hamamatsu R8520-506 photomultiplier tubes were characterized at room temperature and in liquid argon to evaluate their response and operating stability for liquid-argon light readout.

\textbf{Methods} A 405~nm pulsed light source was used to measure gain, peak-to-valley ratio, single-photoelectron charge resolution, and afterpulsing, while dark-count data were acquired with random triggers in the absence of active illumination. 

\textbf{Results} Gain and dark-count stability were monitored during a ten-day operating period in liquid argon. At a motherboard input voltage of 900~V, the extracted gains in liquid argon were approximately 27.8\%--51.3\% higher than the corresponding room-temperature values. For all seven tubes, the peak-to-valley ratios were higher and the single-photoelectron charge resolution improved in liquid argon. The dark count rates ranged from approximately 143 to 252~Hz at room temperature and from 166 to 199~Hz in liquid argon, with three tubes showing increases and four showing decreases. The mean number of detected afterpulses per selected primary event was approximately 21.0\%--63.6\% lower in liquid argon. The relative standard deviations across the ten daily means of gain and dark count rate did not exceed approximately 2.1\% and 3.3\%, respectively. 

\textbf{Conclusion} The measurements provide a quantitative basis for considering R8520-506 tubes as photosensor candidates for liquid-argon detectors.}


\maketitle

\section{Introduction}\label{sec1}

Noble-liquid time projection chambers are widely used in low-energy particle-physics experiments, including dark-matter searches~\cite{Chepel2013,AprileDoke2010}. Energy depositions in liquid argon or liquid xenon generate complementary scintillation and ionization signals for energy measurement, event timing, and position reconstruction. Accurate single-photoelectron (SPE) calibration and stable photosensor gain are therefore important for weak-light measurements, whereas dark counts and afterpulses contribute to optical-readout noise.

Photomultiplier tubes (PMTs) have been developed and qualified for a range of noble-liquid detectors. The R8520 family includes compact metal-package devices such as the R8520-406-001~\cite{HamamatsuR8520406001}. The quantum efficiency of related R8520-406 tubes has been measured at liquid-xenon temperature~\cite{E_Aprile_2012}, while R8520-406 tubes were characterized for PandaX-I in terms of gain, SPE response, random pulse rates, afterpulsing, and operational stability~\cite{Li2016PandaX}. The R11065 family was evaluated at liquid-argon temperature~\cite{Acciarri2012R11065} and subsequently used for the optical readout of DarkSide-50~\cite{Canci2020DS50}. Larger R11410 tubes have undergone dedicated cryogenic-xenon studies~\cite{Baudis2013R11410}, and R11410-21 tubes were qualified in a large-batch campaign for XENON1T~\cite{Barrow_2017}. These studies establish the use of PMTs of different formats for noble-liquid light readout.

The Hamamatsu R8520-506 is a compact one-inch PMT~\cite{HamamatsuR8520506}. To evaluate its performance for liquid-argon light readout, we characterized seven tubes at room temperature and in liquid argon using a common test system. The measurements covered gain, peak-to-valley (P/V) ratio, SPE resolution, dark count rate (DCR), afterpulse timing and rate, and gain and DCR stability during ten days of continuous operation in liquid argon.

\section{Experimental setups}\label{sec:2}

\subsection{Dark-box testing facility}\label{sec:2.1}

\begin{figure}[htbp]
    \centering
    \includegraphics[width=0.9\textwidth]{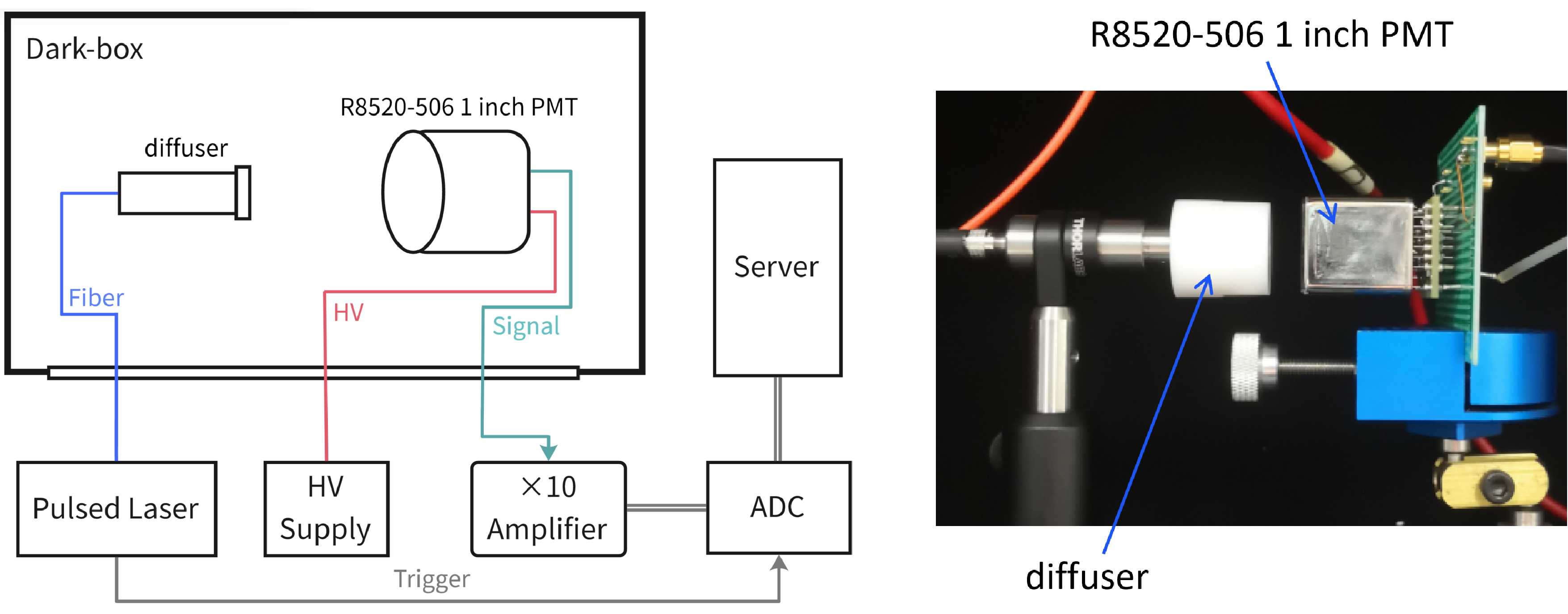}
    \caption{Overview of the dark-box test chain and a photograph of the internal arrangement. The measurements used a Hamamatsu M10306-30 light source with a C10196 controller at a wavelength of 405~nm. In the photograph, the lower and upper blue arrows indicate the diffuser and PMT, respectively.}
    \label{fig:PMT_testing_system_roomT}
\end{figure}

The seven PMTs have identification numbers WA0037, WA0038, WA0039, WA0040, WA0041, WA0043, and WA0045. Each tube was first examined individually in the dark box shown in Figure~\ref{fig:PMT_testing_system_roomT}. Optical pulses at 405~nm were provided by a Hamamatsu M10306-30 light source operated with a C10196 controller and delivered to the photocathode through an optical fiber and diffuser. The PMT anode signal was amplified by a nominal factor of ten before digitization and waveform analysis. For SPE and afterpulse measurements, the acquisition trigger was synchronized with the optical pulse.

The individual dark-box measurements were used to verify the waveform acquisition and the SPE response before the tubes were installed as an array. The performance comparisons reported below were obtained with the dewar system.

\subsection{Dewar system and acquisition modes}\label{sec:2.2}

\begin{figure}[htbp]
    \centering
    \includegraphics[width=0.8\textwidth]{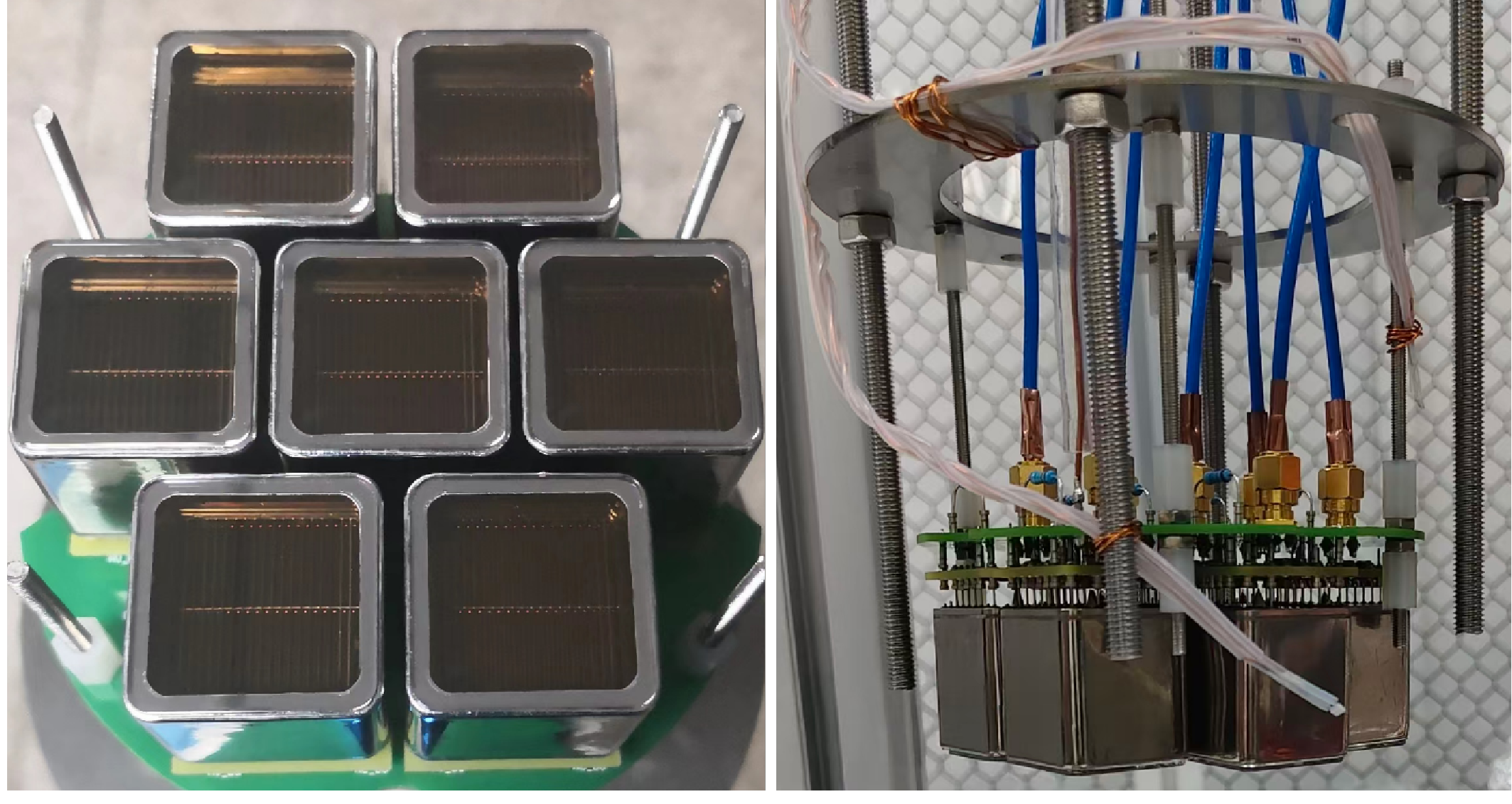}
    \caption{Photographs of the seven-PMT array, the motherboard, and the support structure used for the dewar measurements.}
    \label{fig:R8520_PMT_dewar_motherboard}
\end{figure}

The seven PMTs were mounted facing downward on a common motherboard supported by a metal ring and four threaded rods, as shown in Figure~\ref{fig:R8520_PMT_dewar_motherboard}. A single high-voltage cable supplied the motherboard, which powered the seven PMTs. The same motherboard setting was used for the paired room-temperature and liquid-argon measurements and for the ten-day stability study, with an input voltage of 900~V.

Two acquisition modes were used. Waveforms for SPE calibration and afterpulse analysis were recorded with external triggers synchronized to the 405~nm light pulses. DCR waveforms were acquired using externally generated random triggers and without active illumination. All seven channels were digitized synchronously in the DCR acquisition, but no inter-channel coincidence selection was applied. For the stability study, the PMTs were operated continuously in liquid argon for ten days, while repeated gain and DCR measurements were summarized for each day.

\section{PMT performance measurements and results}\label{sec:3}

The paired room-temperature and liquid-argon measurements were analyzed using the same procedures. The performance comparisons apply to the seven tubes and operating conditions studied here.

\subsection{Gain and single-photoelectron response}\label{sec:3.1}

The waveforms were sampled at 4~ns intervals over a 10~$\mu$s acquisition window, with the external trigger located at 2500~ns. The digitizer output was converted to voltage before waveform analysis. To estimate the baseline, intervals of approximately 50~ns around pre-trigger peaks exceeding approximately 0.1 times the SPE pulse amplitude were excluded from the 0--2200~ns region. The mean of the remaining samples was then subtracted from the full waveform.

The SPE charge was integrated from 2400 to 2800~ns. The integrated baseline-subtracted waveform area was corrected for the nominal tenfold amplifier gain and converted to anode charge as
\begin{equation}
    Q=\frac{1}{A_{\mathrm{amp}}R}
    \int_{2400\,\mathrm{ns}}^{2800\,\mathrm{ns}}V_{\mathrm{b}}(t)\,\mathrm{d}t,
    \label{eq:charge_conversion}
\end{equation}
where $A_{\mathrm{amp}}=10$, $R=50~\Omega$, and $V_{\mathrm{b}}(t)$ is the baseline-subtracted digitizer waveform in volts. The spectrum was expressed as the electron-equivalent charge $q=Q/e$, where $e$ is the elementary charge; consequently, the fitted SPE mean $\mu_1$ gives the PMT gain. The charge spectrum was described by a pedestal Gaussian, a set of $n$-photoelectron Gaussian components, and an exponential component for under-amplified signals~\cite{Barrow_2017}:

\begin{equation}
    f(q)=A_0\exp\left[-\frac{(q-\mu_0)^2}{2\sigma_0^2}\right]
    +\sum_{n=1}^{N_{\max}}A_n\exp\left[-\frac{(q-n\mu_1)^2}{2n\sigma_1^2}\right]
    +B\exp(-q\tau)\,\mathcal{I}_{\mathrm{UA}}(q),
    \label{eq:spe_model}
\end{equation}
Here, $\mu_0$ and $\sigma_0$ describe the pedestal, and $\mu_1$ and $\sigma_1$ are the mean and width of the SPE component in electron-equivalent charge. The $n$-photoelectron components have means $n\mu_1$ and widths $\sqrt{n}\sigma_1$, while their amplitudes are independently fitted; $N_{\max}$ denotes the highest included photoelectron component. The parameters $B$ and $\tau$ describe the amplitude and inverse-charge slope of the under-amplified component. The indicator $\mathcal{I}_{\mathrm{UA}}(q)$ equals one over the fitted charge interval below the SPE peak and zero elsewhere, restricting this component to the low-charge region.

\begin{figure}[htbp]
    \centering
    \includegraphics[width=0.7\textwidth]{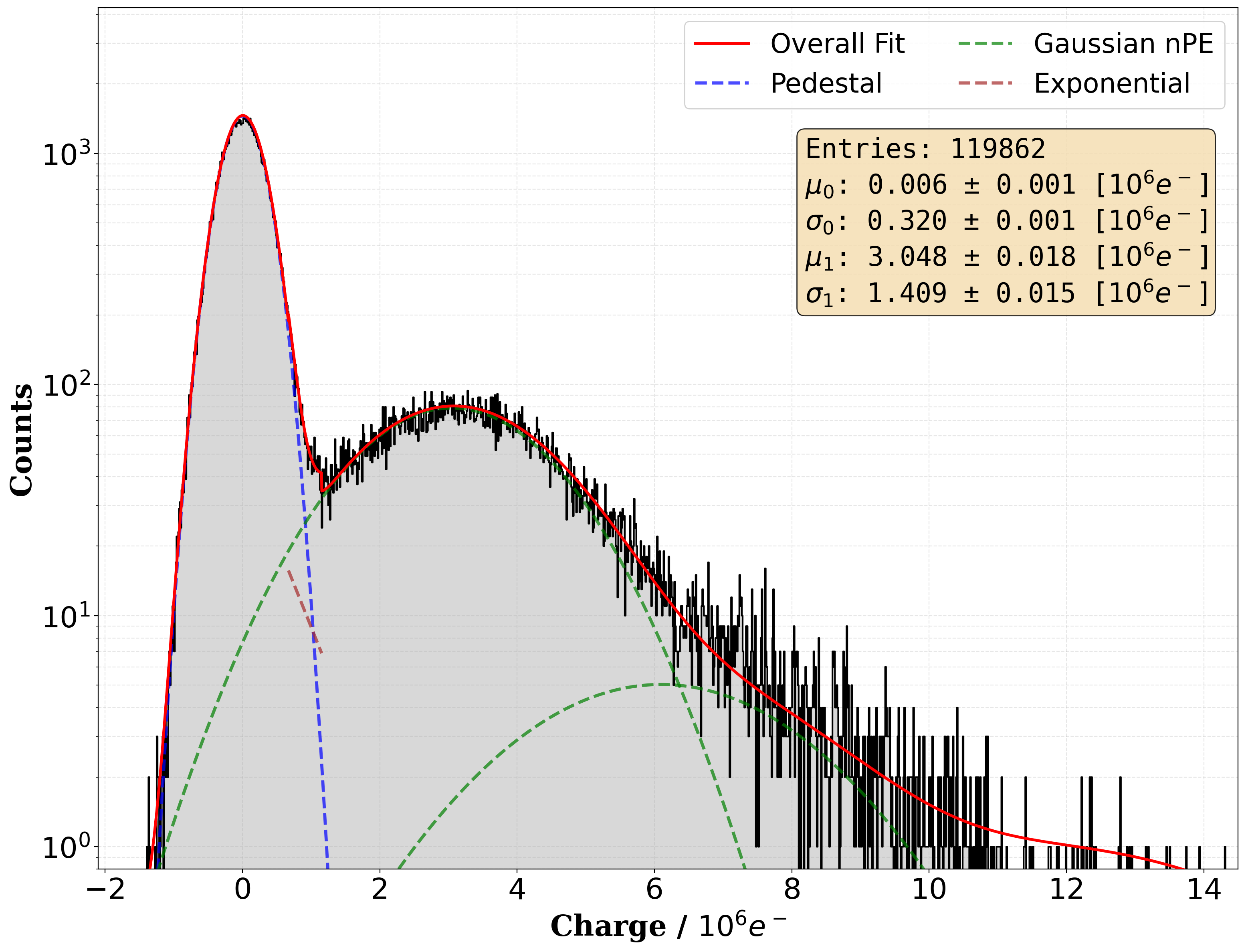}
    \caption{Representative SPE charge spectrum of WA0037 in liquid argon at a motherboard input voltage of 900~V. The charge axis and the Gaussian means and widths shown in the inset are expressed in units of $10^6$ elementary charges. The pedestal, $n$-photoelectron components, under-amplified component, and total fit are shown separately.}
    \label{fig:Fit_spe_spectrum_WA0037_900V}
\end{figure}

Figure~\ref{fig:Fit_spe_spectrum_WA0037_900V} shows a representative fit for WA0037 in liquid argon. The peak-to-valley (P/V) ratio is extracted from the total fitted spectrum as the fitted height at the SPE maximum divided by the fitted minimum between the pedestal and SPE peak. The SPE resolution is defined as $\sigma_1/\mu_1$; a smaller value corresponds to a narrower charge response relative to its mean and hence better SPE charge resolution~\cite{Barrow_2017}.

\begin{figure}[htbp]
    \centering
    \begin{subfigure}[b]{0.48\textwidth}
        \centering
        \includegraphics[width=\textwidth]{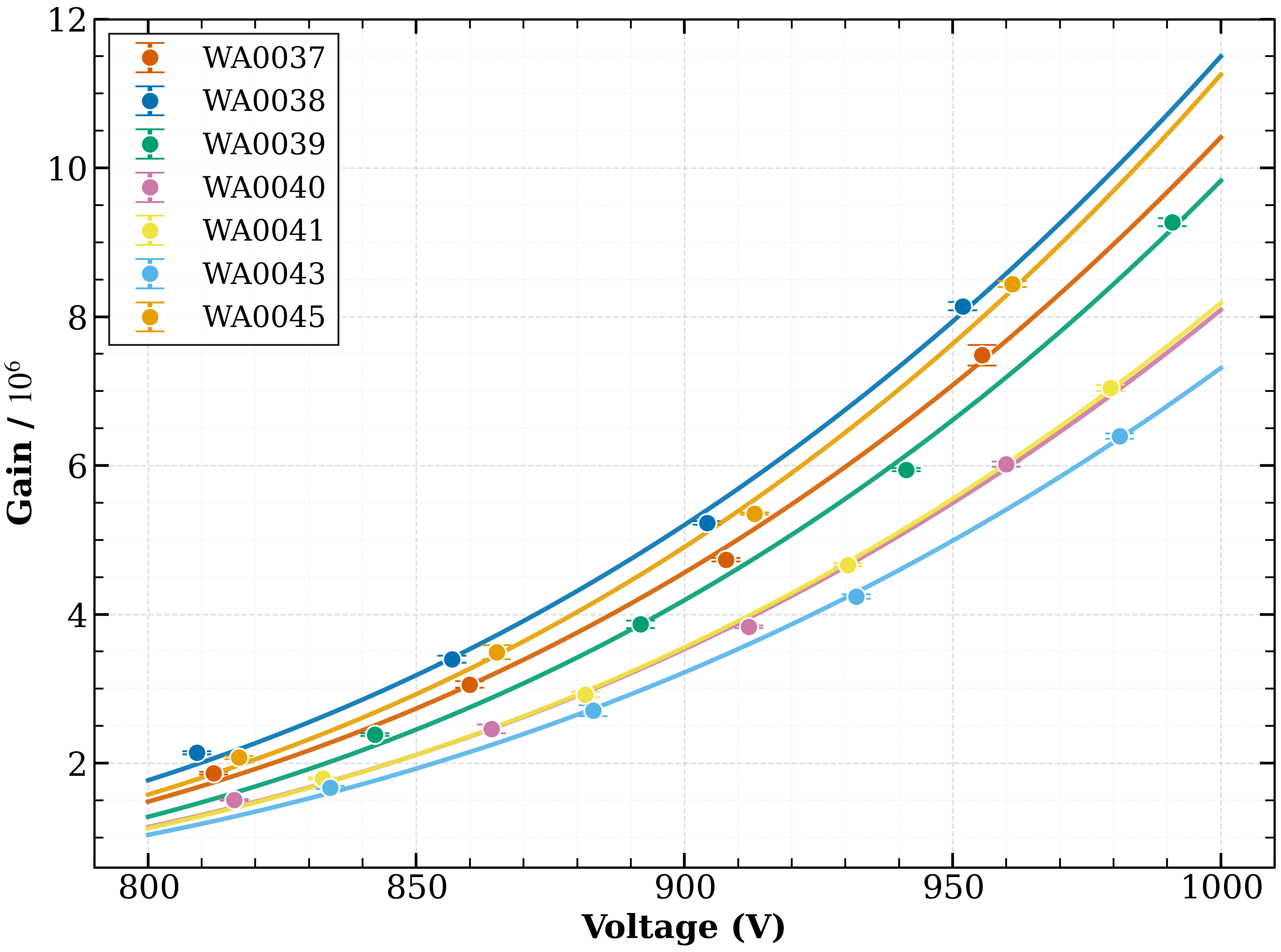}
        \caption{Liquid argon.}
        \label{fig:gain_voltage_lar}
    \end{subfigure}
    \hfill
    \begin{subfigure}[b]{0.48\textwidth}
        \centering
        \includegraphics[width=\textwidth]{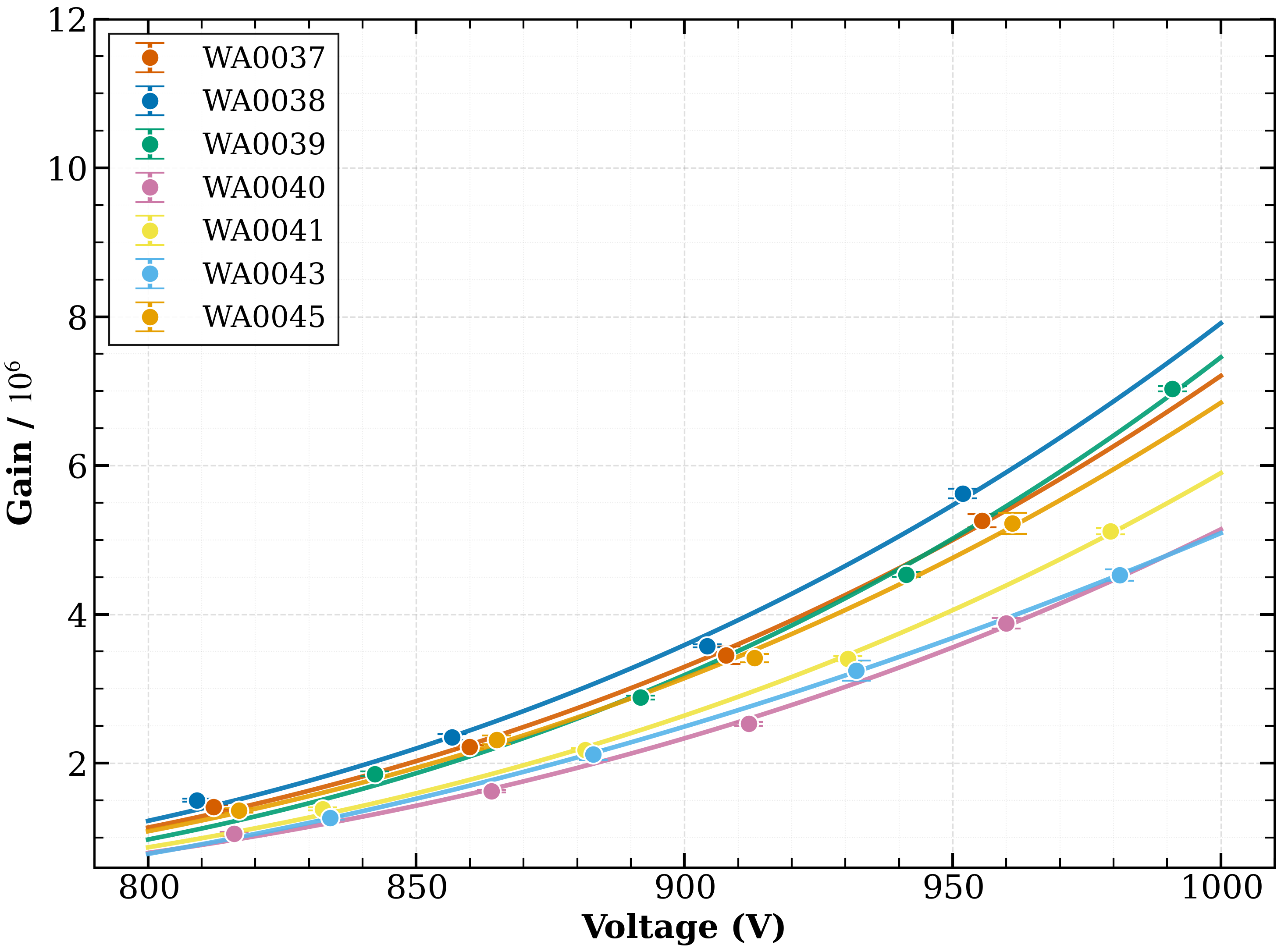}
        \caption{Room temperature.}
        \label{fig:gain_voltage_room}
    \end{subfigure}
    \caption{Gain as a function of applied voltage for the seven PMTs in liquid argon and at room temperature. The curves are empirical fits to the measured points. The error bars are the quadrature sum of the SPE-spectrum fit uncertainty and the statistical standard deviation of repeated measurements.}
    \label{fig:gain_V}
\end{figure}

\begin{figure}[htbp]
    \centering
    \includegraphics[width=0.6\textwidth]{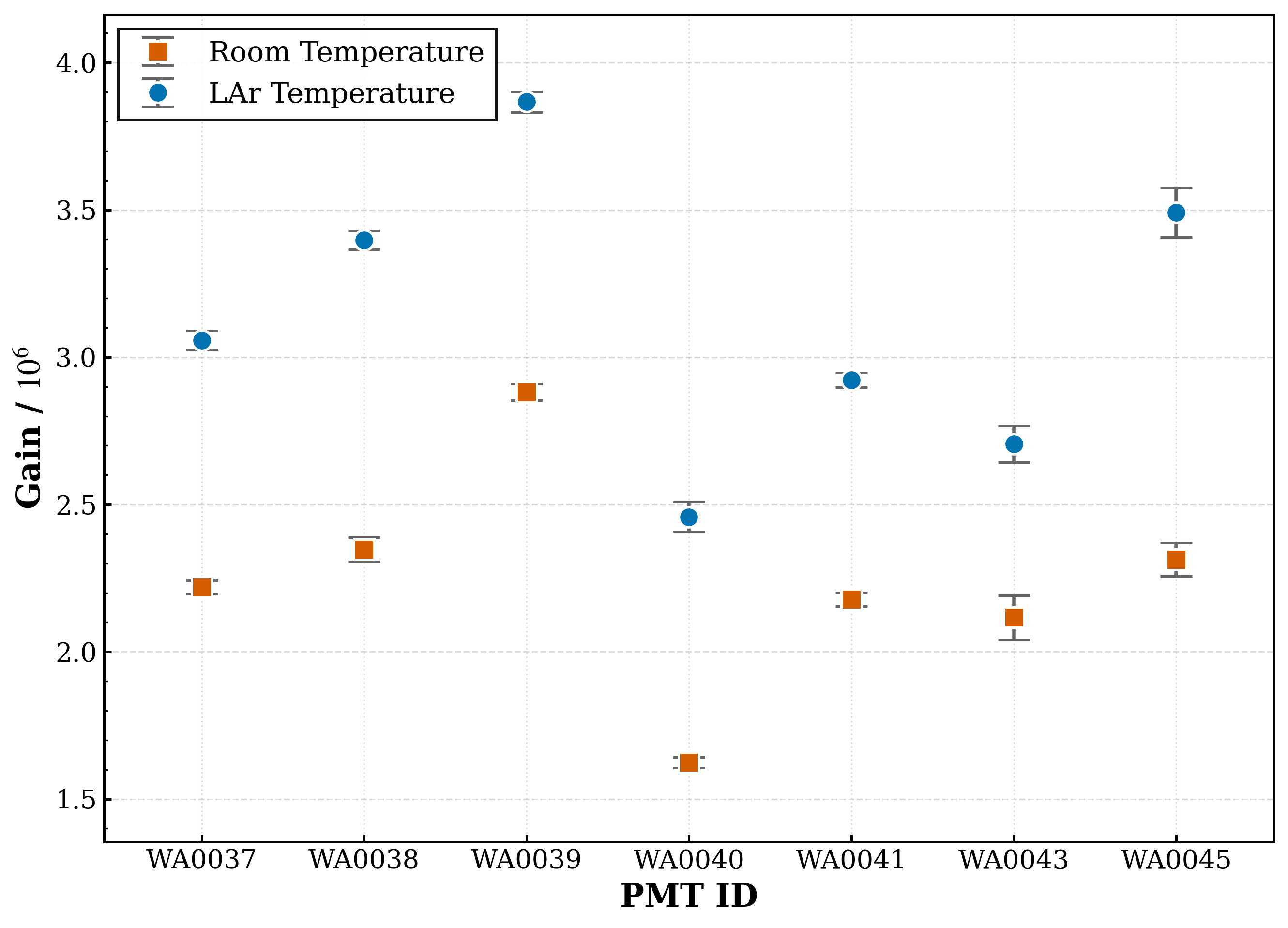}
    \caption{Extracted gains at room temperature and in liquid argon at a motherboard input voltage of 900~V. The error bars are the quadrature sum of the SPE-spectrum fit uncertainty and the statistical standard deviation of repeated measurements.}
    \label{fig:Gain_lar_roomT}
\end{figure}

The gain uncertainty shown in Figures~\ref{fig:gain_V}, \ref{fig:Gain_lar_roomT}, and \ref{fig:Gain_Time_I900_WA0037_45} combines the uncertainty returned by the SPE charge-spectrum fit with the statistical standard deviation of repeated measurements. These contributions are added in quadrature,
\begin{equation}
    \sigma_{G,\mathrm{total}}=
    \sqrt{\sigma_{G,\mathrm{SPE\ fit}}^2+\sigma_{G,\mathrm{stat}}^2}.
    \label{eq:gain_uncertainty}
\end{equation}

Figure~\ref{fig:gain_V} summarizes the measured voltage dependence of the extracted gain at room temperature and in liquid argon. The fitted curves provide an empirical description of the observed dependence.

At a motherboard input voltage of 900~V, the measured gains of all seven PMTs are higher in liquid argon than at room temperature, as shown in Figure~\ref{fig:Gain_lar_roomT}. The relative increases calculated from the plotted values range from approximately 27.8\% to 51.3\%.

\begin{figure}[htbp]
    \centering
    \begin{subfigure}[b]{0.48\textwidth}
        \centering
        \includegraphics[width=\textwidth]{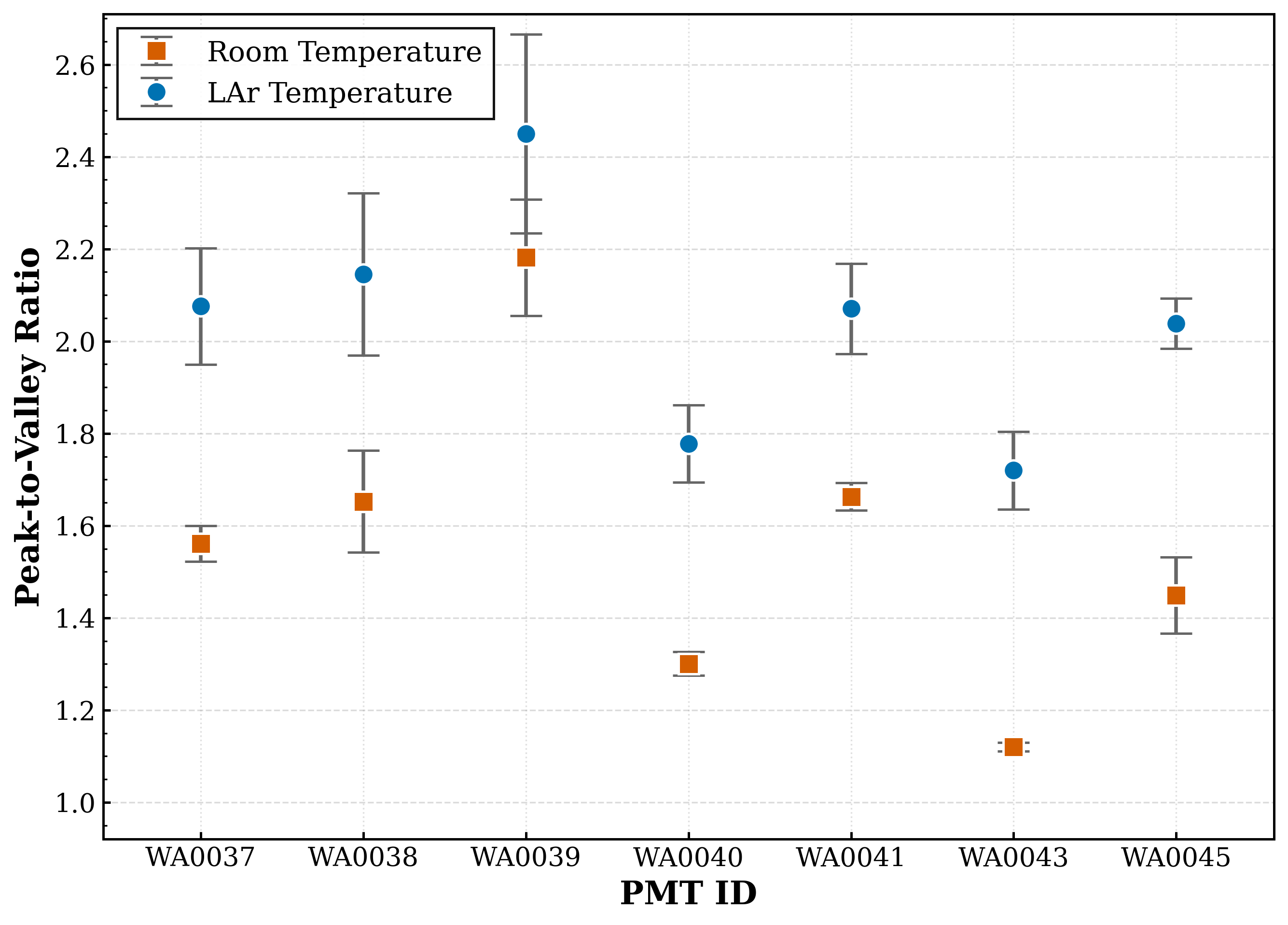}
        \caption{Peak-to-valley ratio.}
        \label{fig:pv_comparison}
    \end{subfigure}
    \hfill
    \begin{subfigure}[b]{0.48\textwidth}
        \centering
        \includegraphics[width=\textwidth]{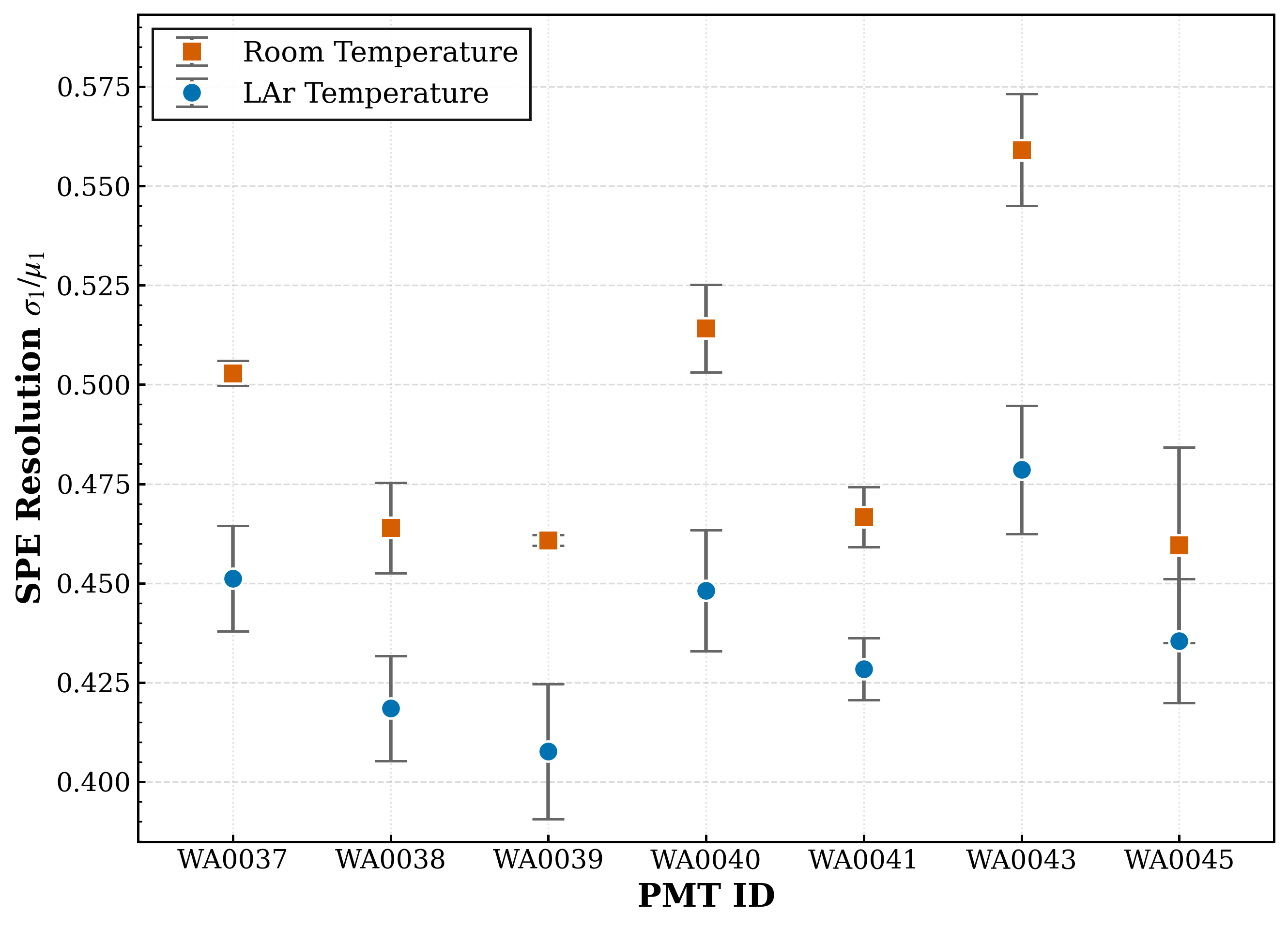}
        \caption{SPE resolution.}
        \label{fig:spe_resolution_comparison}
    \end{subfigure}
    \caption{Comparison of the SPE response at room temperature and in liquid argon at a motherboard input voltage of 900~V. The SPE resolution is defined as $\sigma_1/\mu_1$. The error bars represent the statistical standard deviation of repeated measurements.}
    \label{fig:PVratio_RES_lar_roomT}
\end{figure}

For all seven PMTs, the P/V ratios are higher in liquid argon than at room temperature. The P/V ratios of WA0040 and WA0043 nevertheless remain below 2. The relative SPE charge width, $\sigma_1/\mu_1$, decreases by approximately 5.3\% to 14.4\% in liquid argon, indicating improved SPE charge resolution. These comparisons are shown in Figure~\ref{fig:PVratio_RES_lar_roomT}; their error bars represent the statistical standard deviation of repeated measurements.

\subsection{Dark count rate}\label{sec:3.2}

DCR data were recorded with externally generated random triggers in the absence of active illumination. A pulse was counted when its amplitude exceeded 0.3 times the corresponding SPE pulse amplitude. Assuming a stationary Poisson count process within the acquisition windows, the DCR was obtained from the zero-count fraction~\cite{adrover2025characterization}:
\begin{equation}
    r_{\mathrm{DC}}=-\frac{\ln(N_0/N)}{\Delta t},
    \label{eq:dcr}
\end{equation}
where $N_0$ is the number of windows containing no above-threshold pulse, $N$ is the total number of randomly triggered windows, and $\Delta t=10~\mu$s is the duration of one window. Each data set contained approximately $10^6$ acquisition windows. The DCR was evaluated in two-minute intervals, and the standard deviation of the interval-level values was used to characterize the measurement spread.

The seven channels were analyzed independently. The pulses are interpreted as PMT-internal dark emission, with thermionic emission as a possible contribution; the random-trigger measurement does not distinguish the underlying emission mechanisms.

\begin{figure}[htbp]
    \centering
    \includegraphics[width=0.6\textwidth]{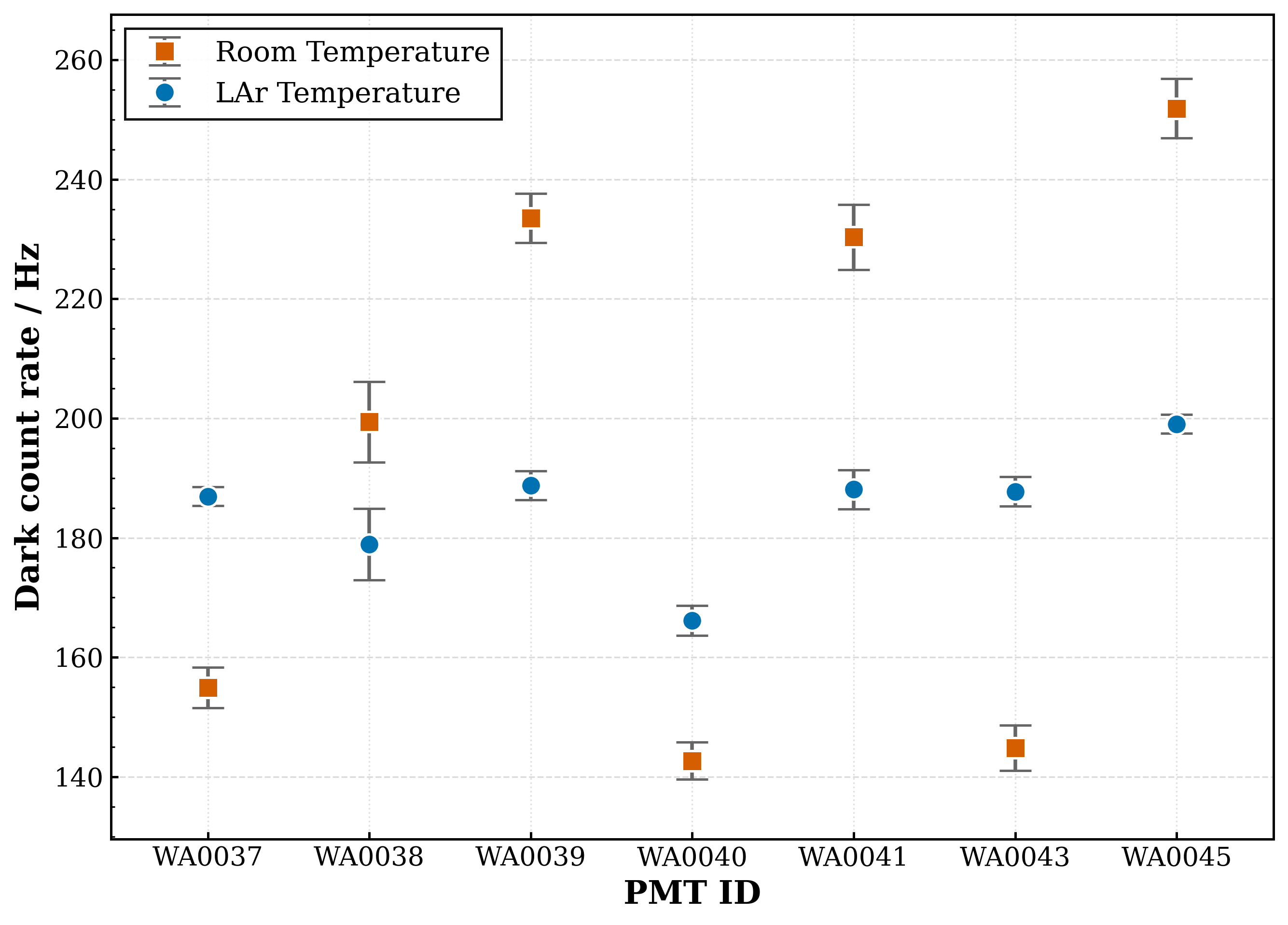}
    \caption{DCRs of the seven PMTs at room temperature and in liquid argon at a motherboard input voltage of 900~V, obtained from randomly triggered acquisition windows without active illumination. The error bars represent the statistical standard deviation of repeated measurements.}
    \label{fig:DCR_lar_roomT}
\end{figure}

Figure~\ref{fig:DCR_lar_roomT} shows that the DCRs span approximately 143--252~Hz at room temperature and 166--199~Hz in liquid argon. The DCR values increase for three tubes and decrease for four, so the seven-tube sample does not exhibit a common directional change between the two temperature conditions.

\subsection{Afterpulse characteristics}\label{sec:3.3}

Fast afterpulses may arise from electron backscattering at the first dynode, whereas slow afterpulses are mainly attributed to residual-gas ionization and subsequent ion feedback~\cite{Barrow_2017}. In the latter process, positive ions produced by the primary electron avalanche drift back toward the photocathode and generate delayed secondary emission. For a fixed production region and electric-field configuration, the ion drift time is expected to scale approximately with $\sqrt{m_i/q_i}$, where $m_i/q_i$ is the ion mass-to-charge ratio. Because the internal field and ion-production positions are not available for the tested tubes, this relation is used only to guide qualitative candidate assignments.

The afterpulse analysis used laser-triggered 405~nm data. Within each record, a primary light pulse was identified when a peak at the trigger position exceeded 0.3 times the SPE pulse amplitude. Only events satisfying this primary-pulse criterion entered the denominator; records without a qualifying primary peak were excluded. For these selected events, a 6000~ns afterpulse search window began when the primary waveform fell below 1\% of its peak amplitude. Each peak above 0.3 times the SPE pulse amplitude in this window contributed to the numerator, allowing multiple counts from a single event. The afterpulse rate is defined as
\begin{equation}
    R_{\mathrm{AP}}=
    \frac{N_{\mathrm{AP\ peaks}}}{N_{\mathrm{selected\ primary\ events}}},
    \label{eq:afterpulse_rate}
\end{equation}
where $N_{\mathrm{AP\ peaks}}$ is the number of afterpulse peaks in the selected events and $N_{\mathrm{selected\ primary\ events}}$ includes all events satisfying the primary-pulse criterion, even if no afterpulse is detected. Thus, $R_{\mathrm{AP}}$ is the mean number of detected afterpulses per selected primary event, not the fraction of events containing an afterpulse.

\begin{figure}[htbp]
    \centering
    \begin{subfigure}[b]{0.49\textwidth}
        \centering
        \includegraphics[width=\textwidth]{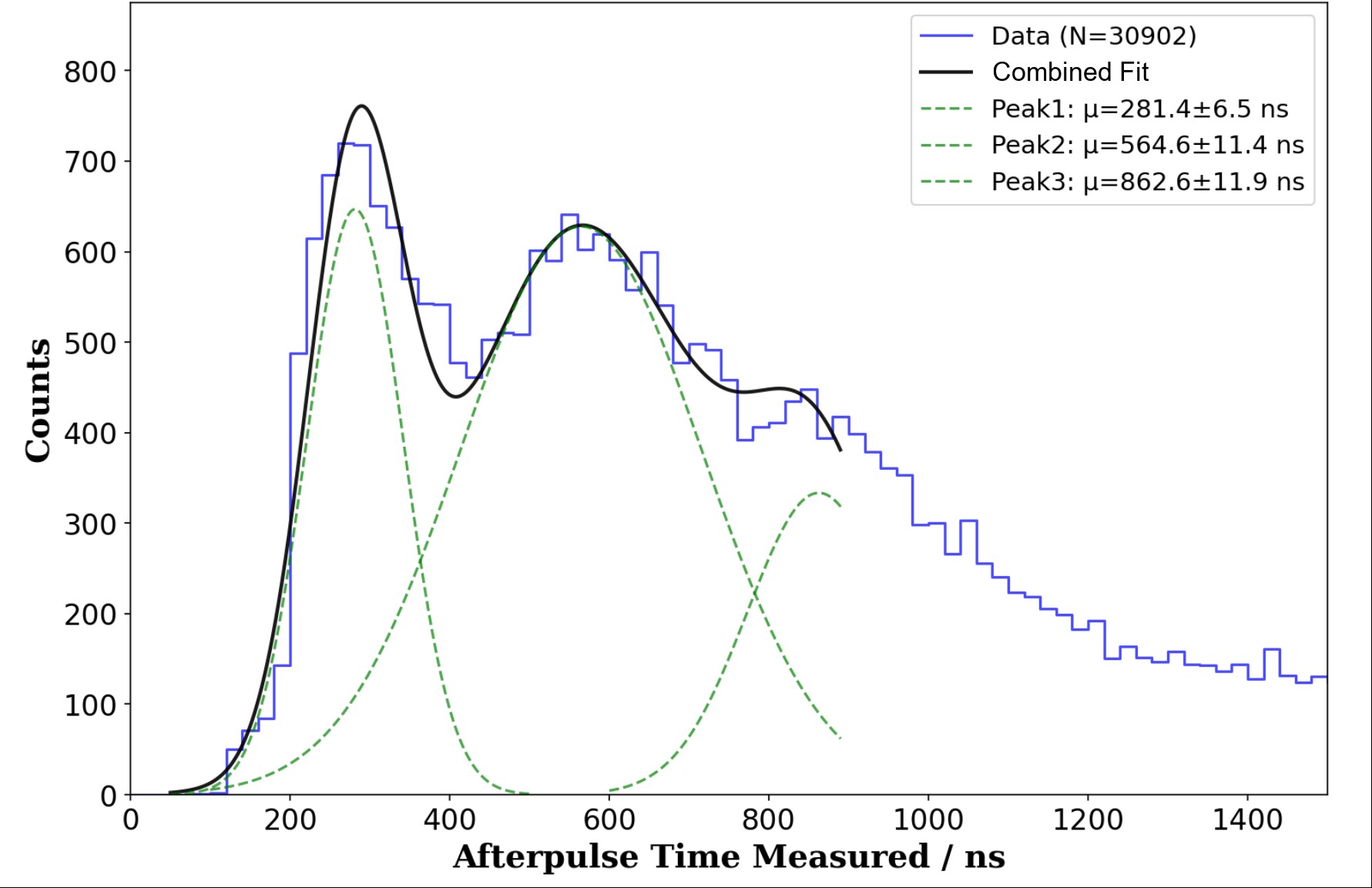}
        \caption{Liquid argon.}
    \end{subfigure}
    \hfill
    \begin{subfigure}[b]{0.49\textwidth}
        \centering
        \includegraphics[width=\textwidth]{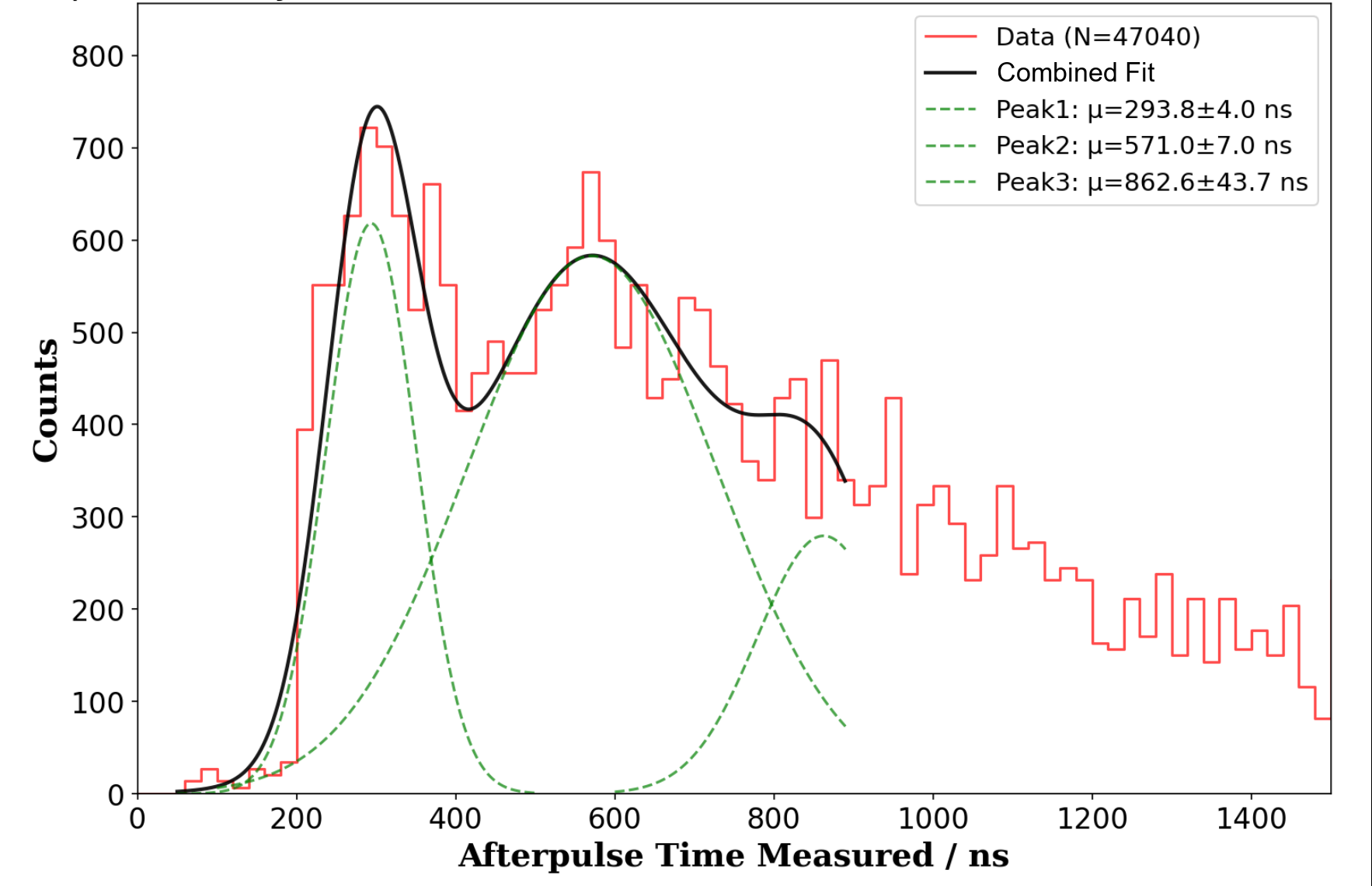}
        \caption{Room temperature.}
    \end{subfigure}
    \caption{Raw-count distributions of the slow-afterpulse delays for WA0037, with the selection $t>50$~ns applied. The delay $t$ is measured from the primary-pulse maximum. The short-delay population at $t<50$~ns is not shown. The dashed green Gaussian components indicate the three fitted timing structures, and the black curve is the combined fit. The candidate ion assignments discussed in the text are not unique identifications.}
    \label{fig:AP_versus_mark}
\end{figure}

Figure~\ref{fig:AP_versus_mark} presents the raw-count distributions of the afterpulse delay $t$ for WA0037, measured relative to the primary-pulse maximum. A $t>50$~ns selection is applied to these distributions to focus on the slow-afterpulse component. The excluded $t<50$~ns region has substantially higher counts than the later-delay region and may be dominated by fast afterpulses from electron backscattering~\cite{Barrow_2017}. This time selection applies only to the two timing distributions and is not used in the afterpulse-rate calculation defined in Equation~\ref{eq:afterpulse_rate}.

Three timing structures in the slow-afterpulse distributions were fitted with Gaussian components. Their centers are $281.4\pm6.5$, $564.6\pm11.4$, and $862.6\pm11.9$~ns in liquid argon, compared with $293.8\pm4.0$, $571.0\pm7.0$, and $862.6\pm43.7$~ns at room temperature; the quoted uncertainties are fit uncertainties. The relative locations are compatible with He$^+$, CH$_4^+$, and Ar$^+$ as candidate interpretations under the simplified $\sqrt{m_i/q_i}$ scaling, but the timing data do not uniquely identify the ion species. The distributions are not normalized to the number of selected primary events, so their histogram heights are not used to compare afterpulse rates.

\begin{figure}[htbp]
    \centering
    \includegraphics[width=0.6\textwidth]{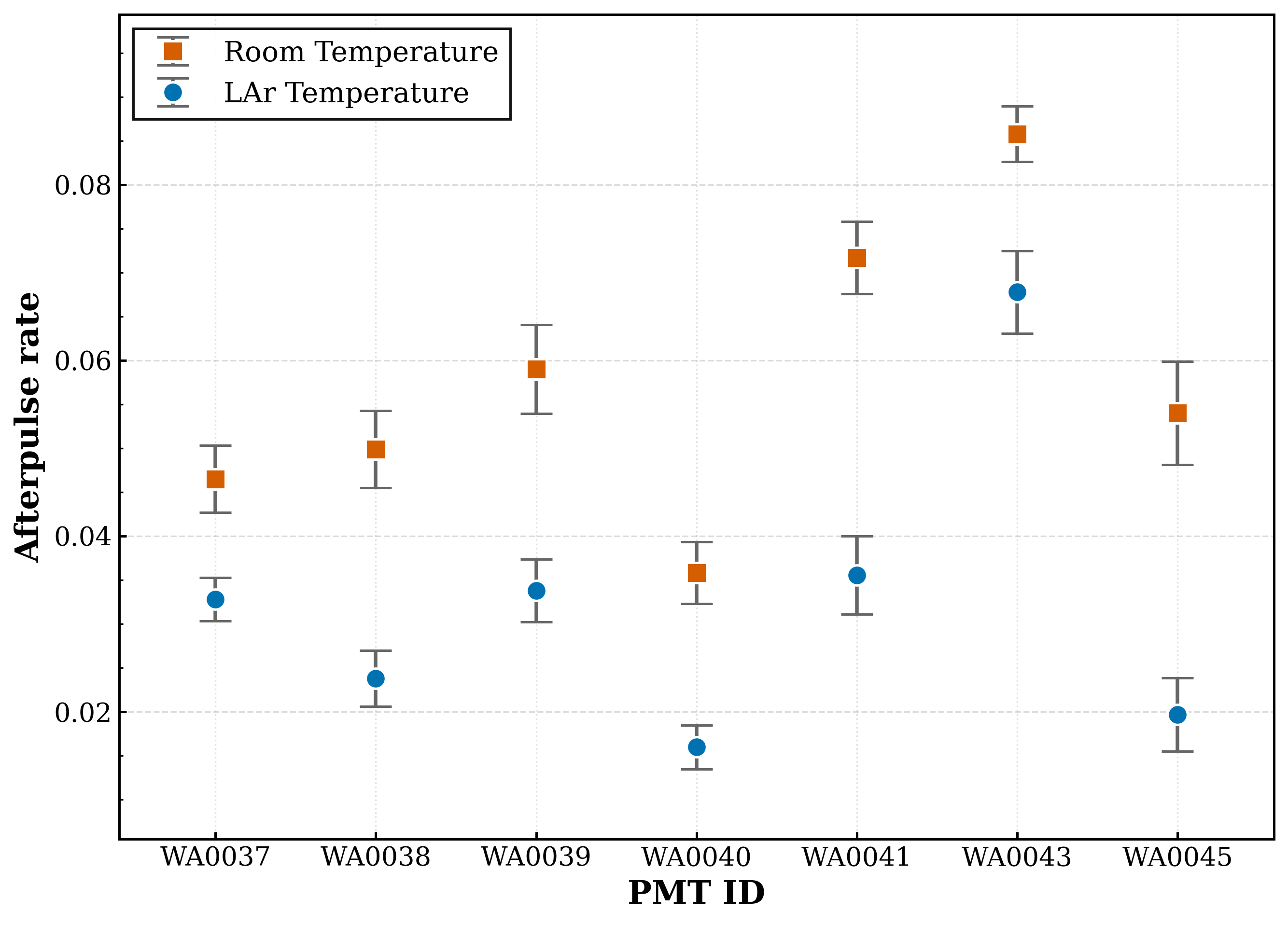}
    \caption{Afterpulse rates at room temperature and in liquid argon at a motherboard input voltage of 900~V. The rate is the mean number of detected afterpulse peaks per laser-triggered event satisfying the primary-pulse criterion. The error bars represent the statistical standard deviation of repeated measurements.}
    \label{fig:APratio_lar_roomT}
\end{figure}

The afterpulse rates of all seven PMTs are lower in liquid argon than at room temperature, as shown in Figure~\ref{fig:APratio_lar_roomT}. The decreases calculated from the plotted values range from approximately 21.0\% to 63.6\%. This comparison quantifies the measured change but does not identify its microscopic origin.

\subsection{Long-term stability over a ten-day operating period}\label{sec:3.4}

\begin{figure}[htbp]
    
    \begin{minipage}[b]{0.47\textwidth}
        \centering
        \includegraphics[width=\textwidth]{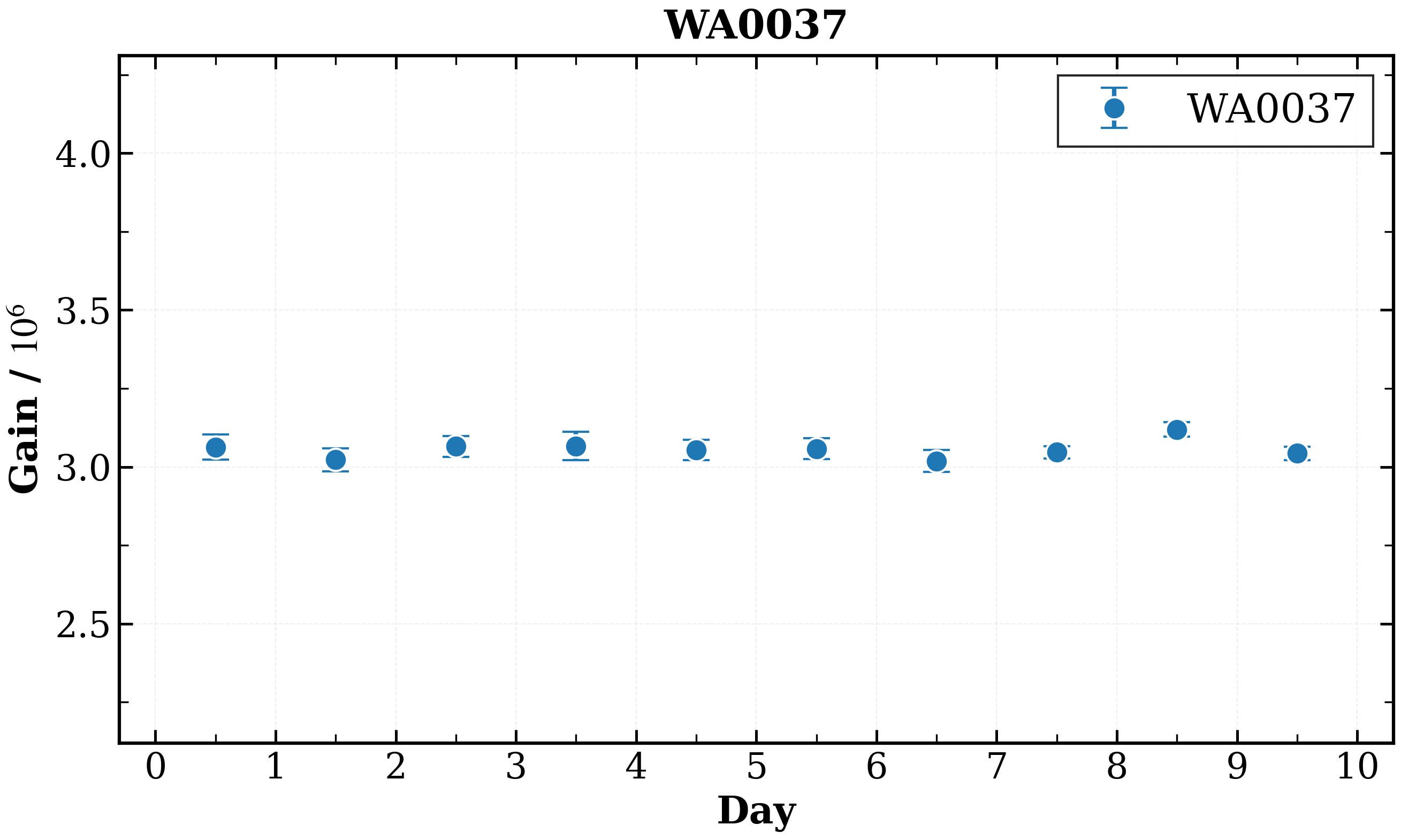}
    \end{minipage}
    \hspace{0.03\textwidth}
    \begin{minipage}[b]{0.47\textwidth}
        \centering
        \includegraphics[width=\textwidth]{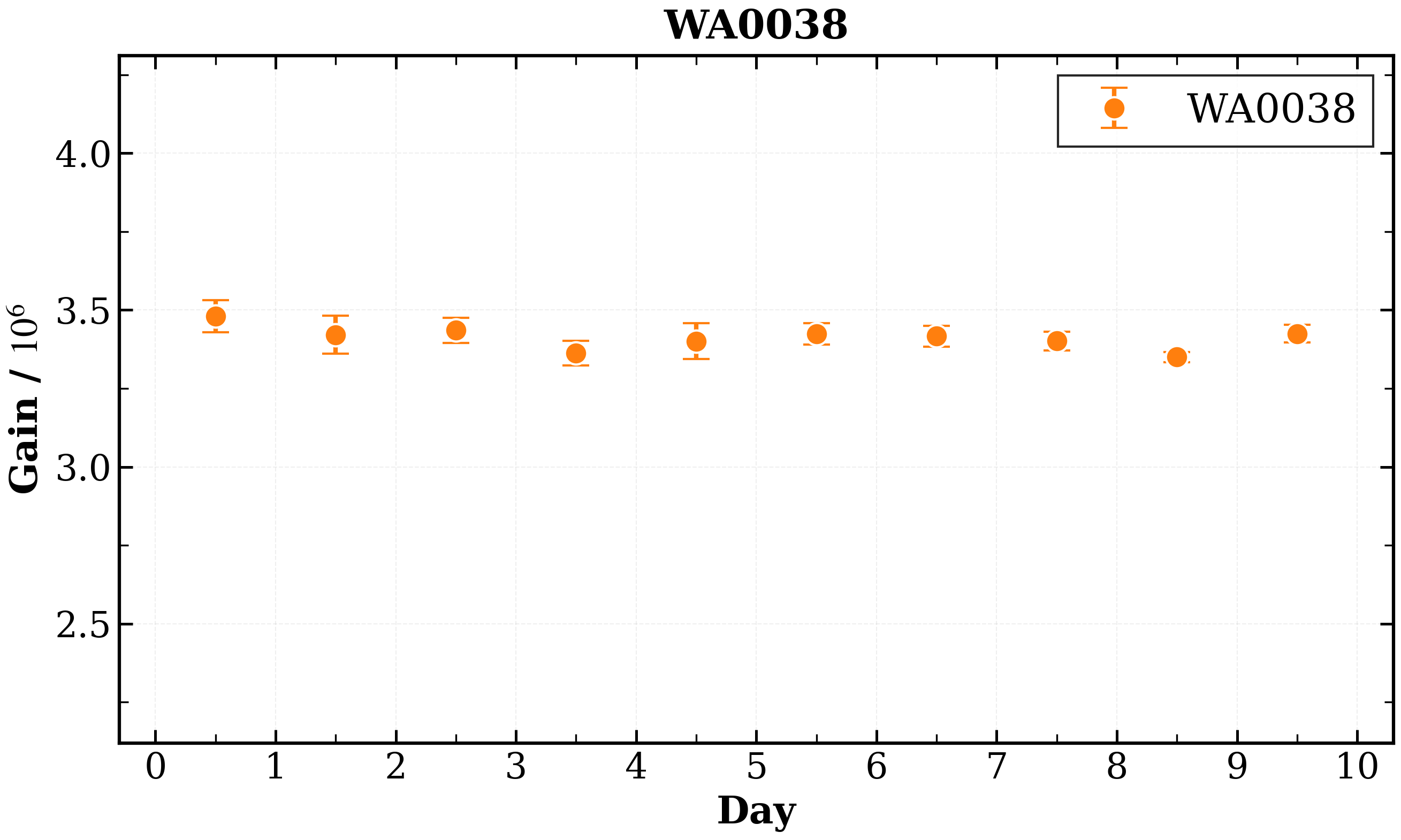}
    \end{minipage}
    \\[-0.7em]
    
    \begin{minipage}[b]{0.47\textwidth}
        \centering
        \includegraphics[width=\textwidth]{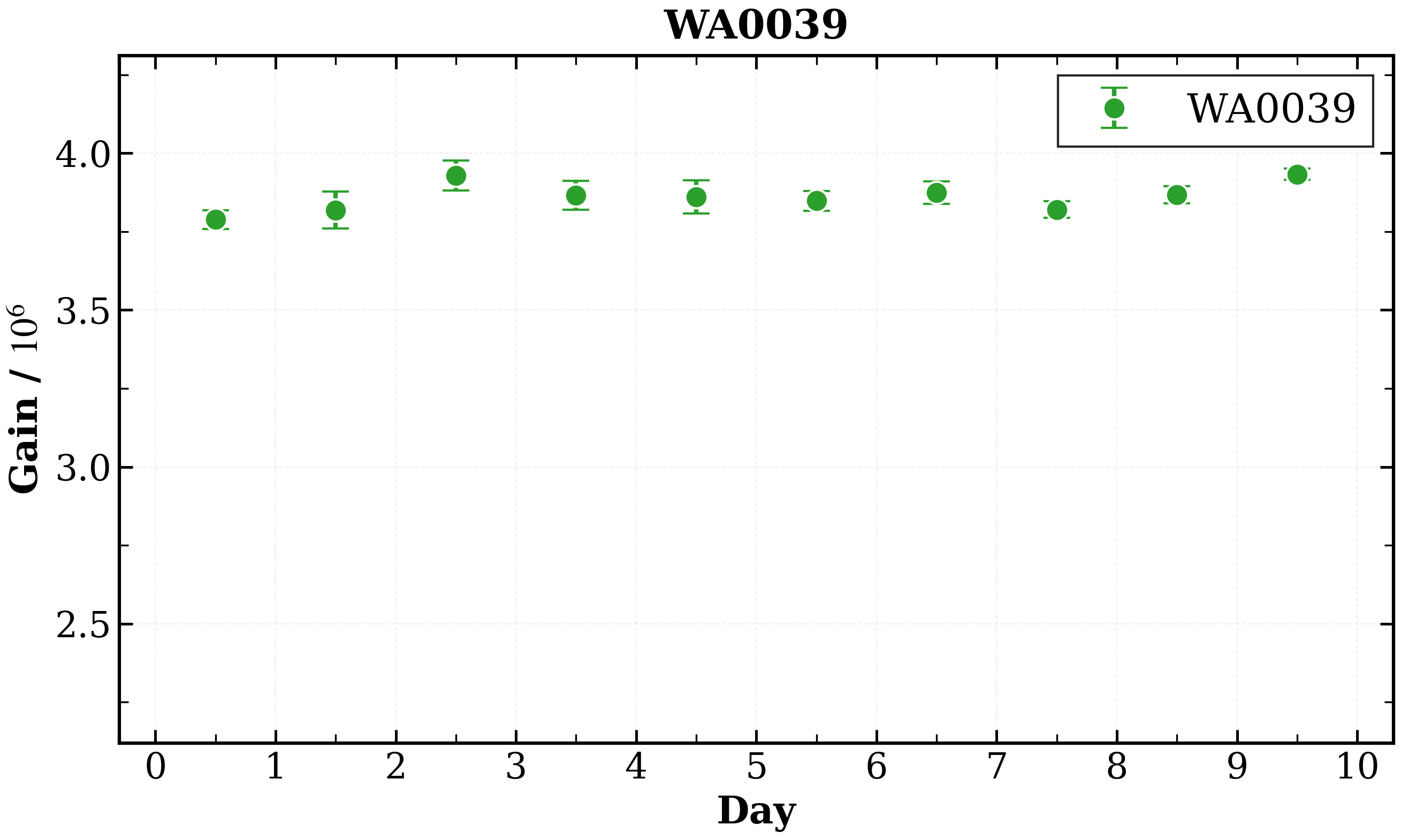}
    \end{minipage}
    \hspace{0.03\textwidth}
    \begin{minipage}[b]{0.47\textwidth}
        \centering
        \includegraphics[width=\textwidth]{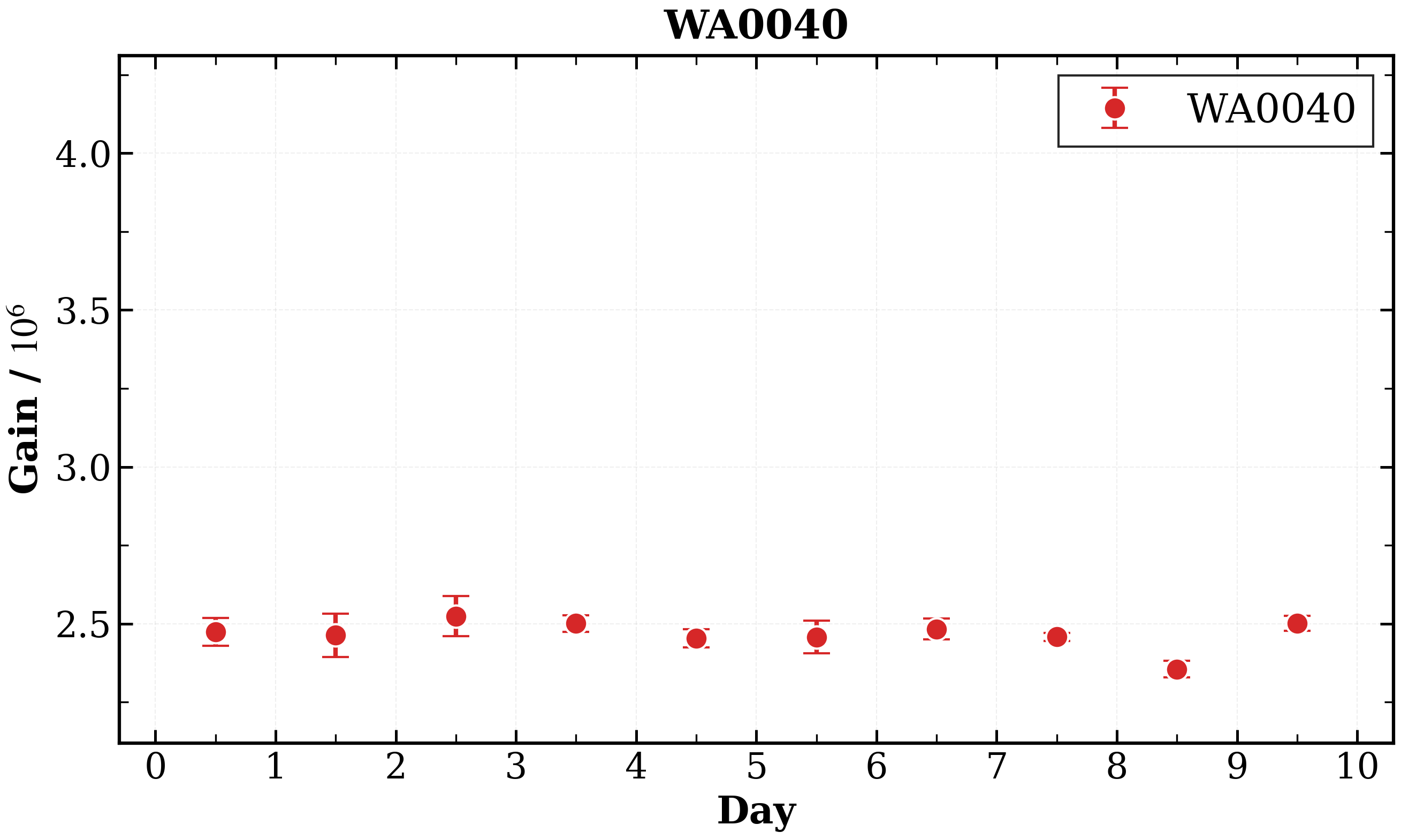}
    \end{minipage}
    \\[-0.7em]
    
    \begin{minipage}[b]{0.47\textwidth}
        \centering
        \includegraphics[width=\textwidth]{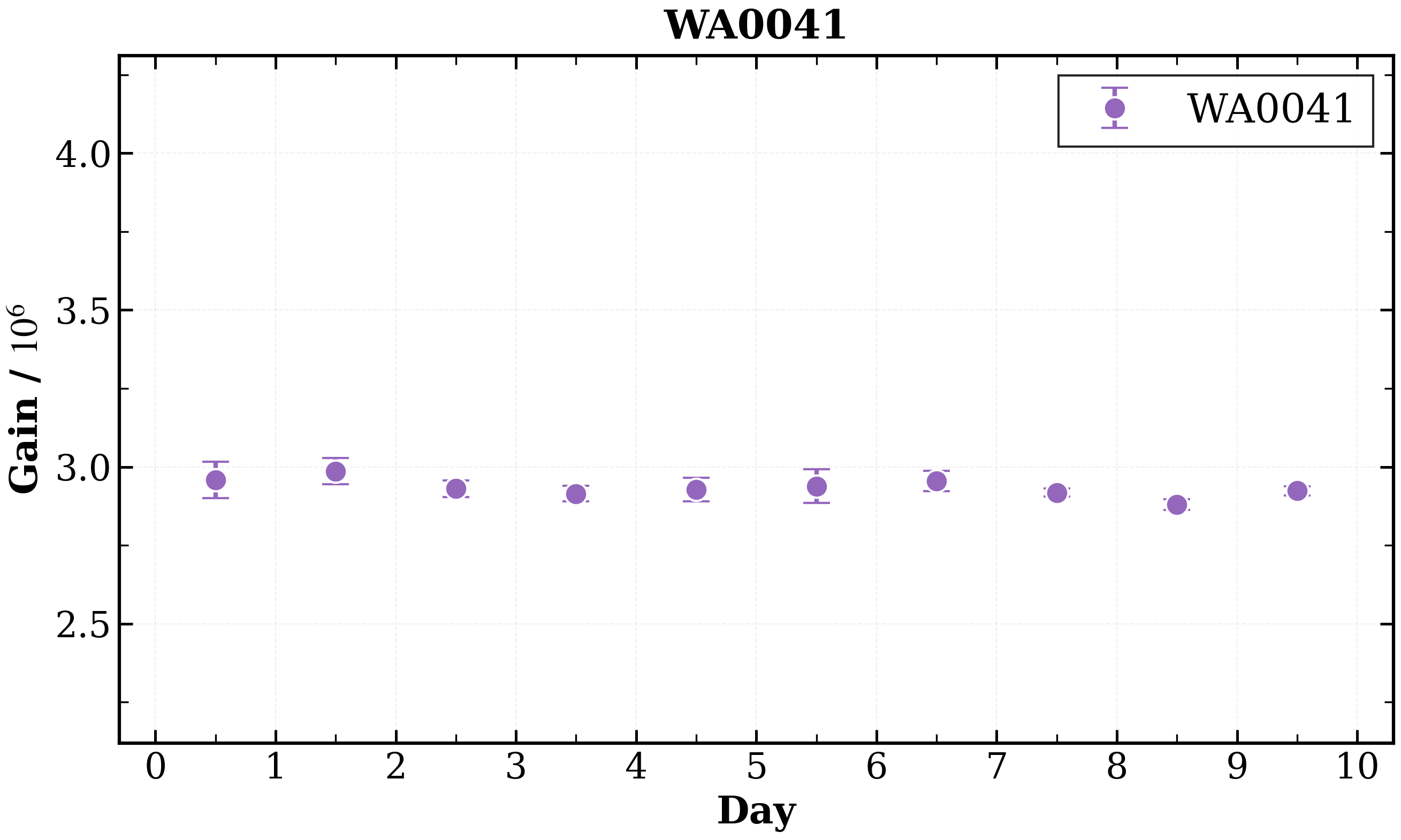}
    \end{minipage}
    \hspace{0.03\textwidth}
    \begin{minipage}[b]{0.47\textwidth}
        \centering
        \includegraphics[width=\textwidth]{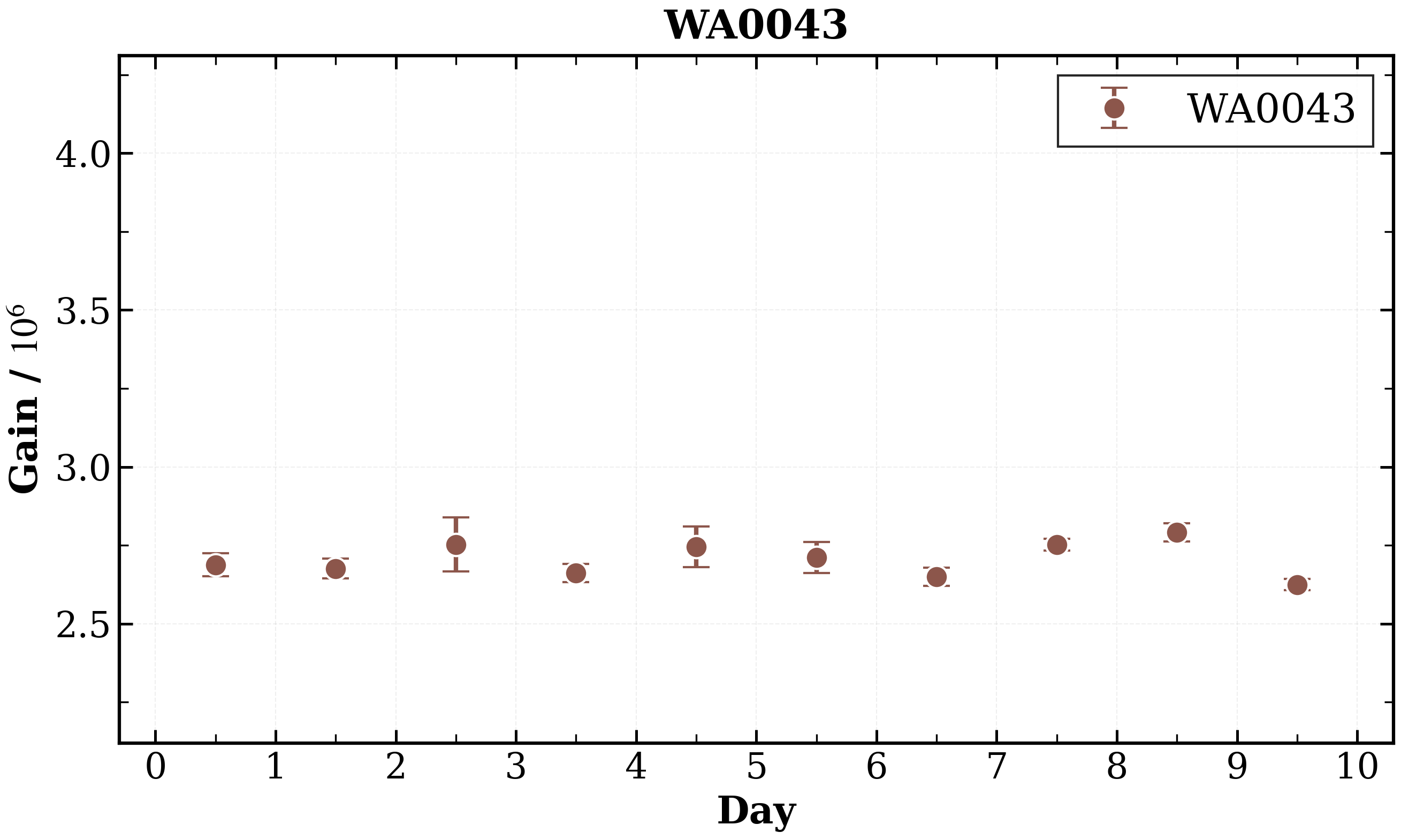}
    \end{minipage}
    \\[-0.7em]
    
    \begin{minipage}[b]{0.47\textwidth}
        \includegraphics[width=\textwidth]{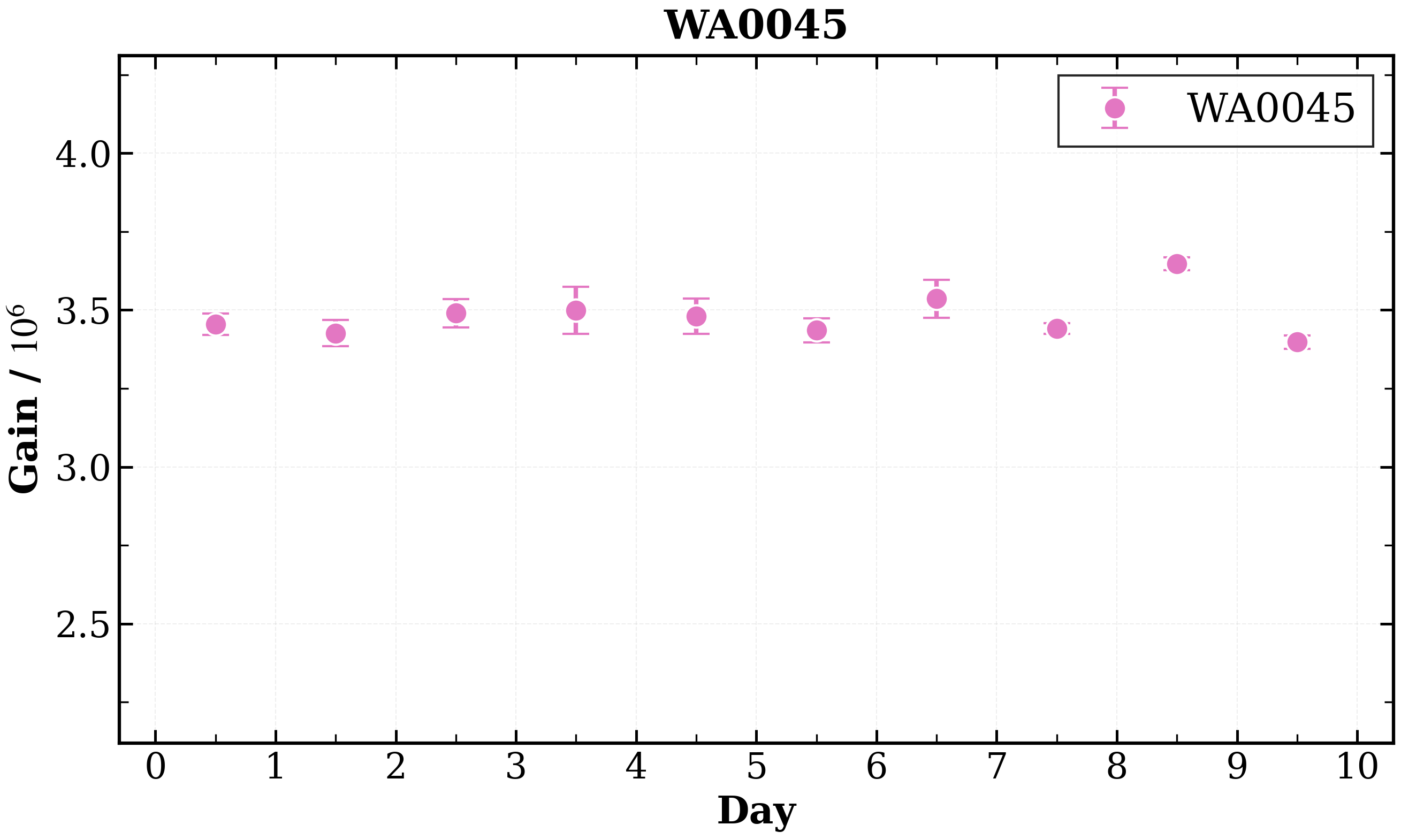}
    \end{minipage}
    
    \caption{Daily mean gains during the ten-day operating period in liquid argon at a motherboard input voltage of 900~V. Each point is plotted at the daily midpoint. The error bars are the quadrature sum of the SPE-spectrum fit uncertainty and the statistical standard deviation of repeated measurements within that day.}
    \label{fig:Gain_Time_I900_WA0037_45}
\end{figure}

\begin{figure}[htbp]
    
    \begin{minipage}[b]{0.47\textwidth}
        \centering
        \includegraphics[width=\textwidth]{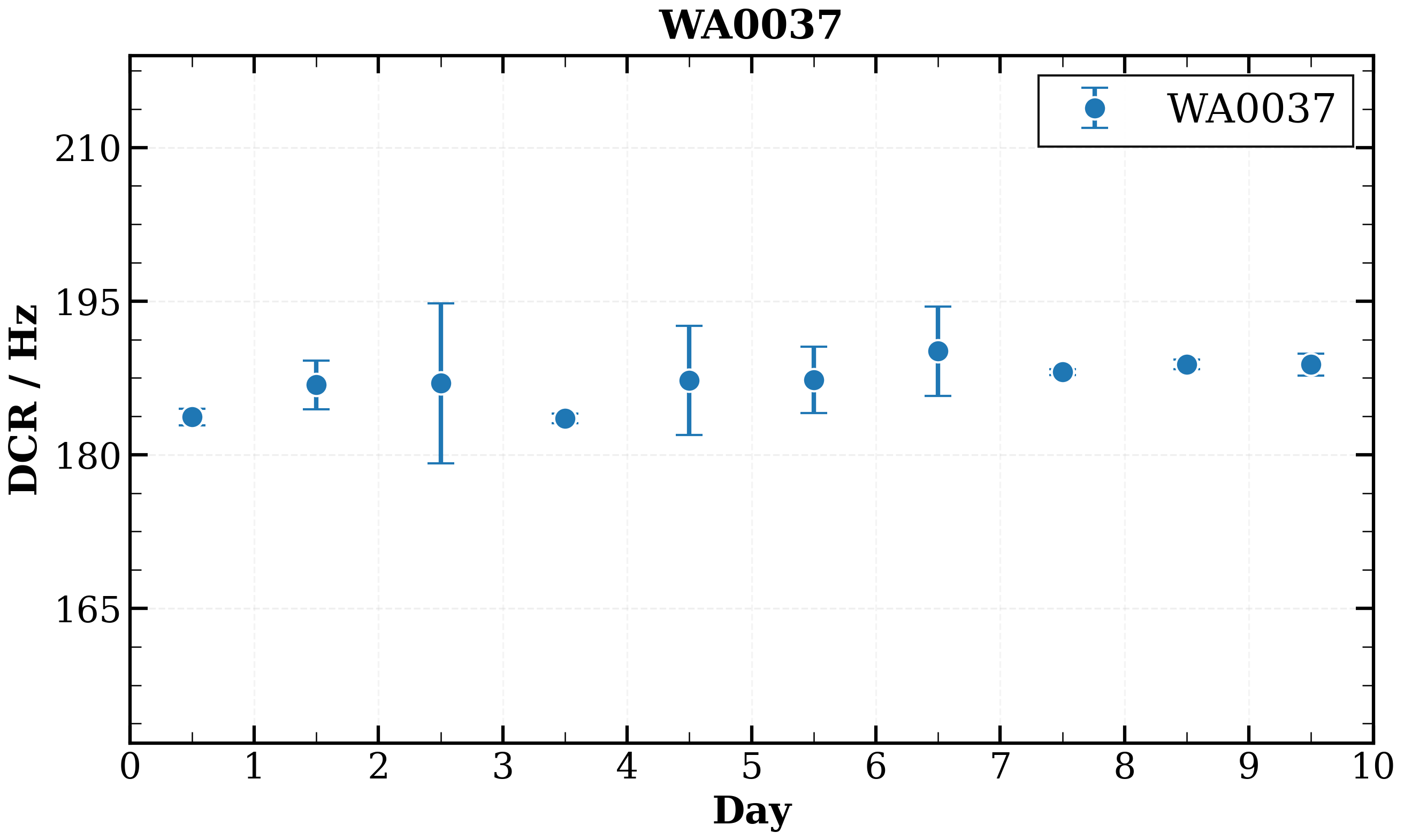}
    \end{minipage}
    \hspace{0.03\textwidth}
    \begin{minipage}[b]{0.47\textwidth}
        \centering
        \includegraphics[width=\textwidth]{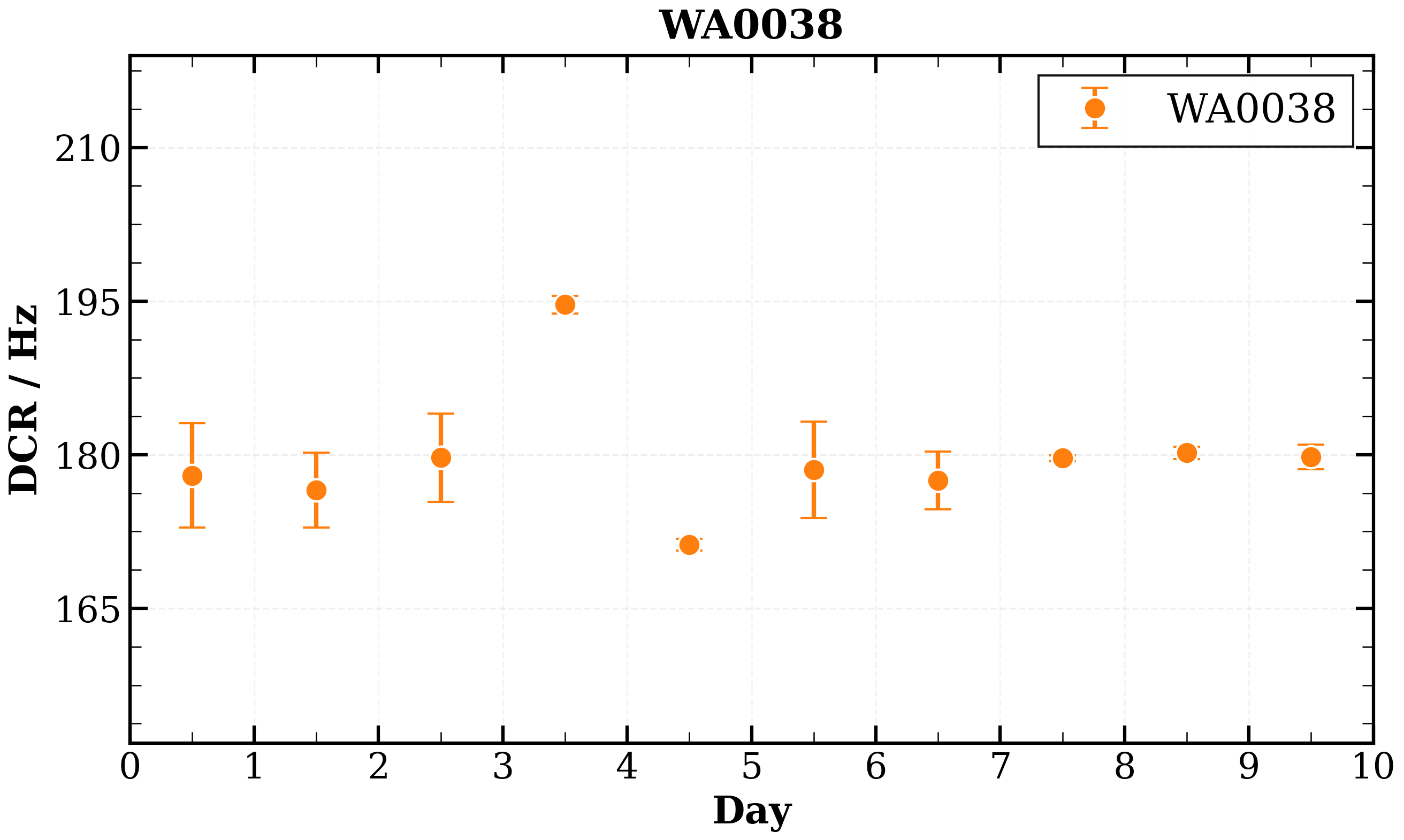}
    \end{minipage}
    \\[-0.7em]
    
    \begin{minipage}[b]{0.47\textwidth}
        \centering
        \includegraphics[width=\textwidth]{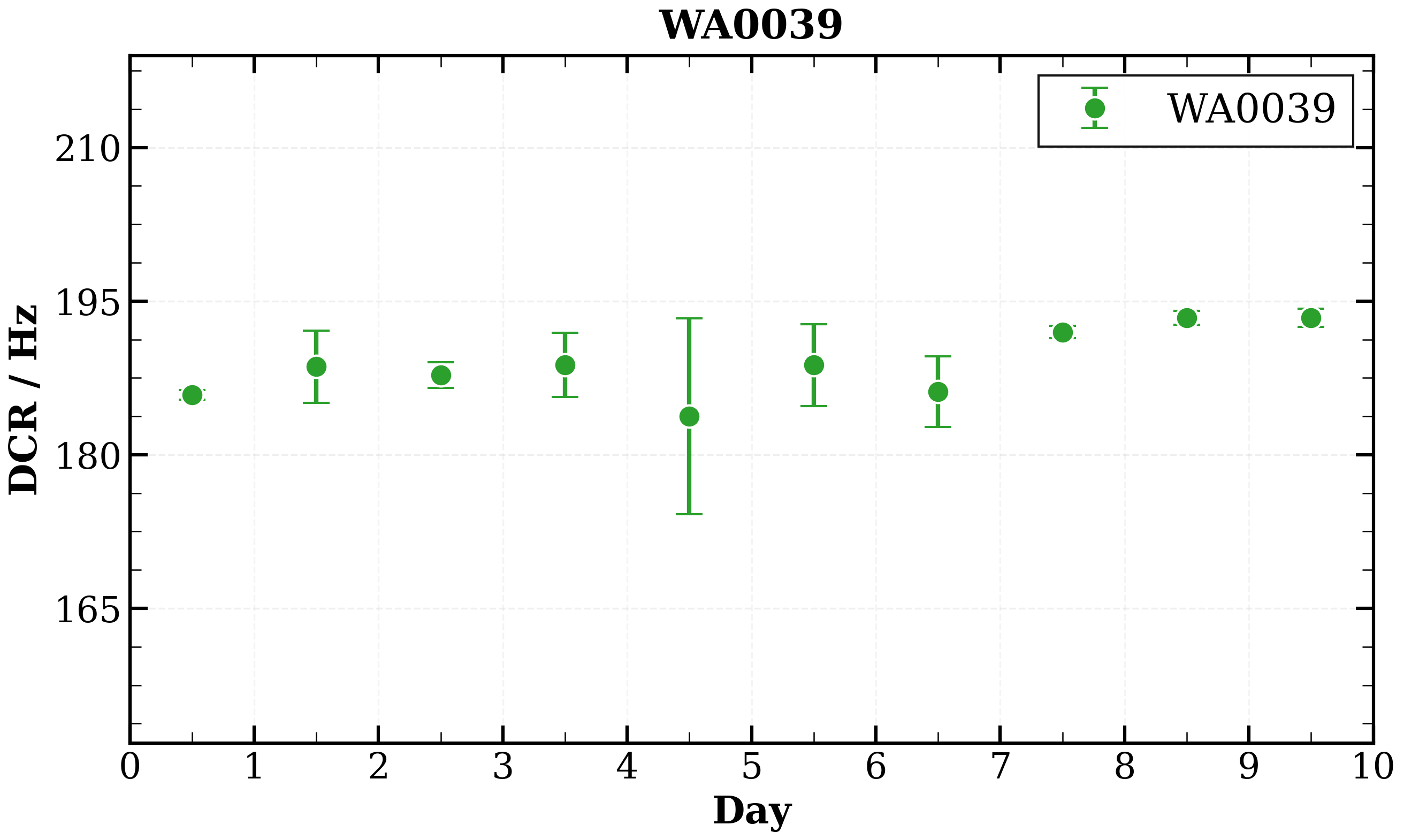}
    \end{minipage}
    \hspace{0.03\textwidth}
    \begin{minipage}[b]{0.47\textwidth}
        \centering
        \includegraphics[width=\textwidth]{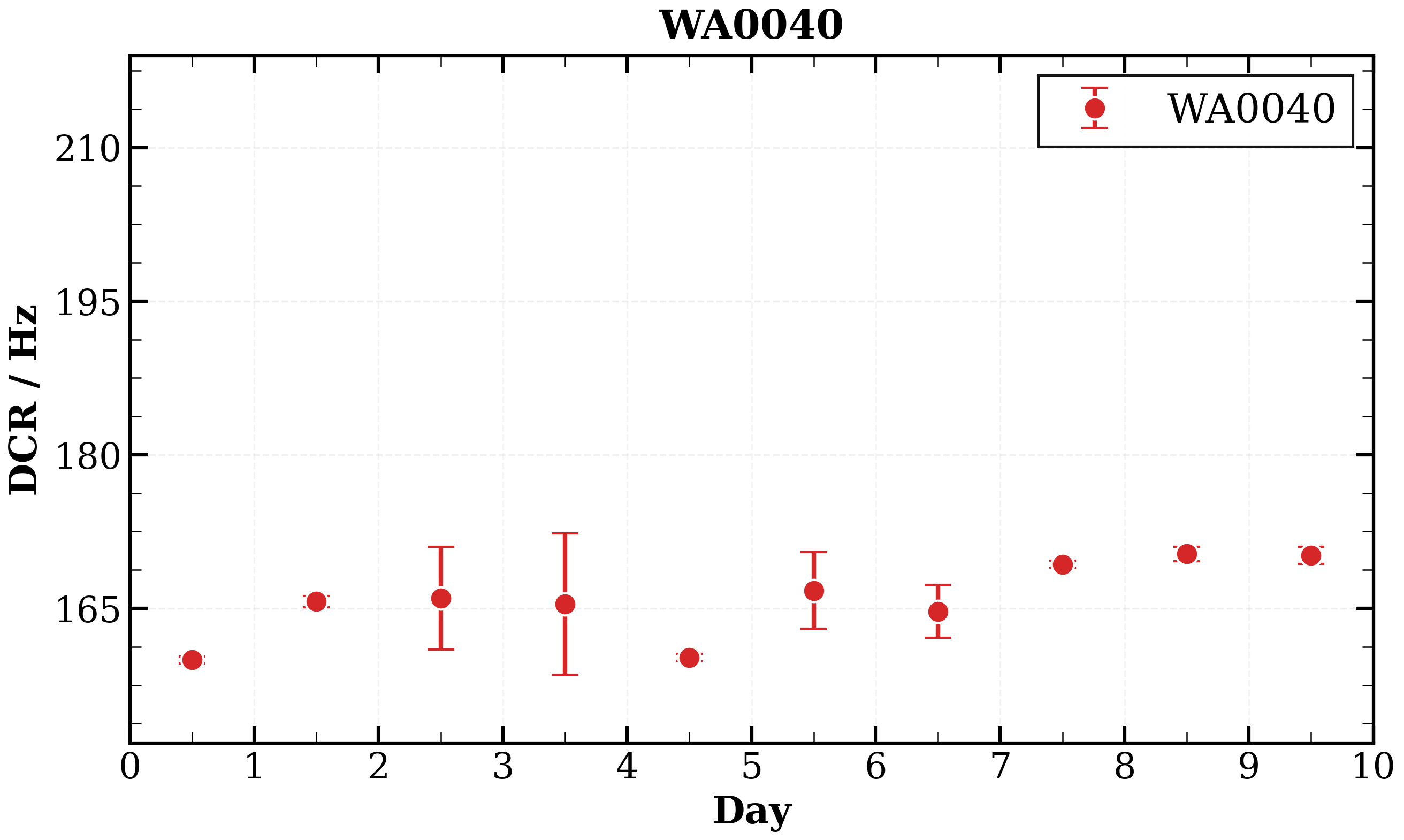}
    \end{minipage}
    \\[-0.7em]
    
    \begin{minipage}[b]{0.47\textwidth}
        \centering
        \includegraphics[width=\textwidth]{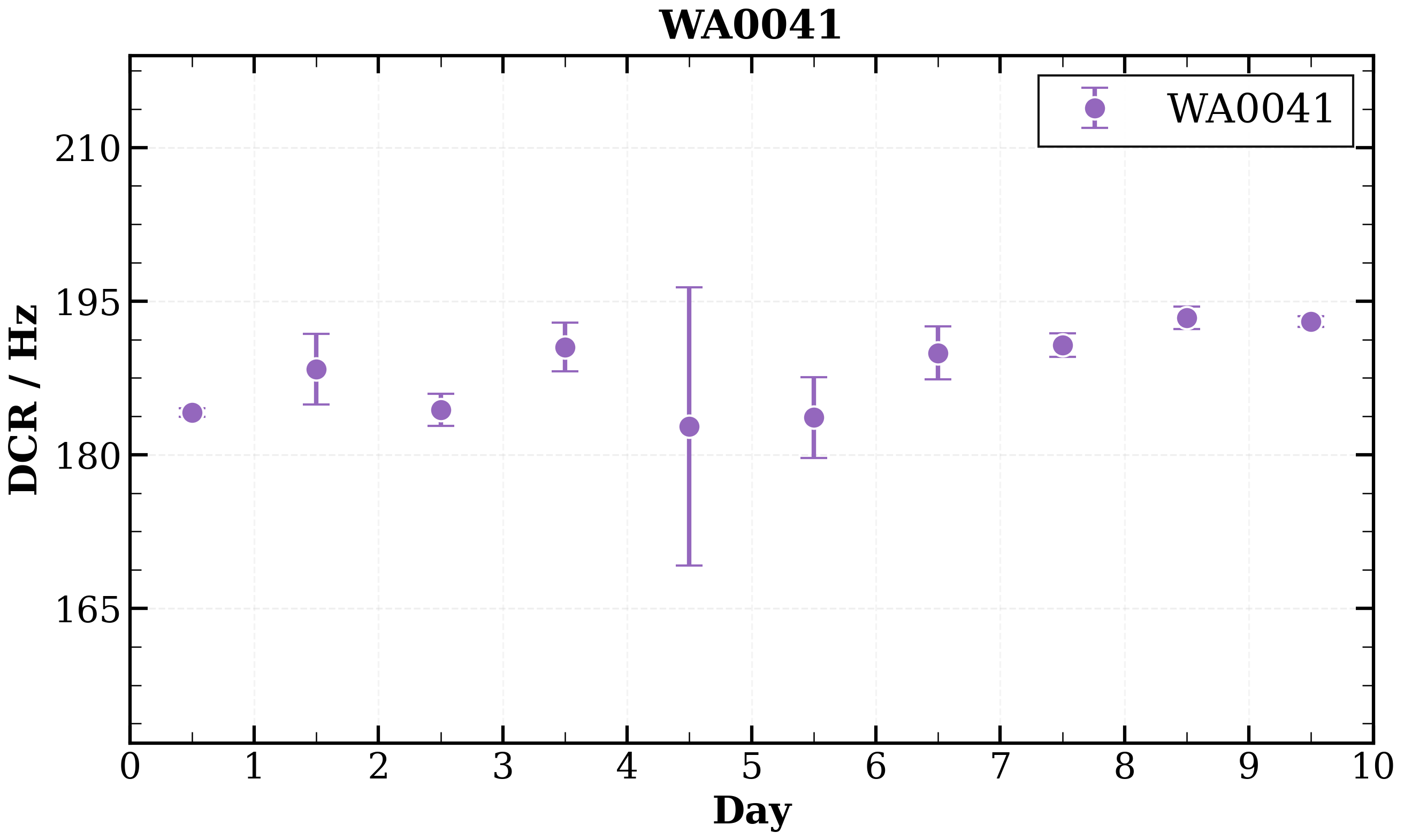}
    \end{minipage}
    \hspace{0.03\textwidth}
    \begin{minipage}[b]{0.47\textwidth}
        \centering
        \includegraphics[width=\textwidth]{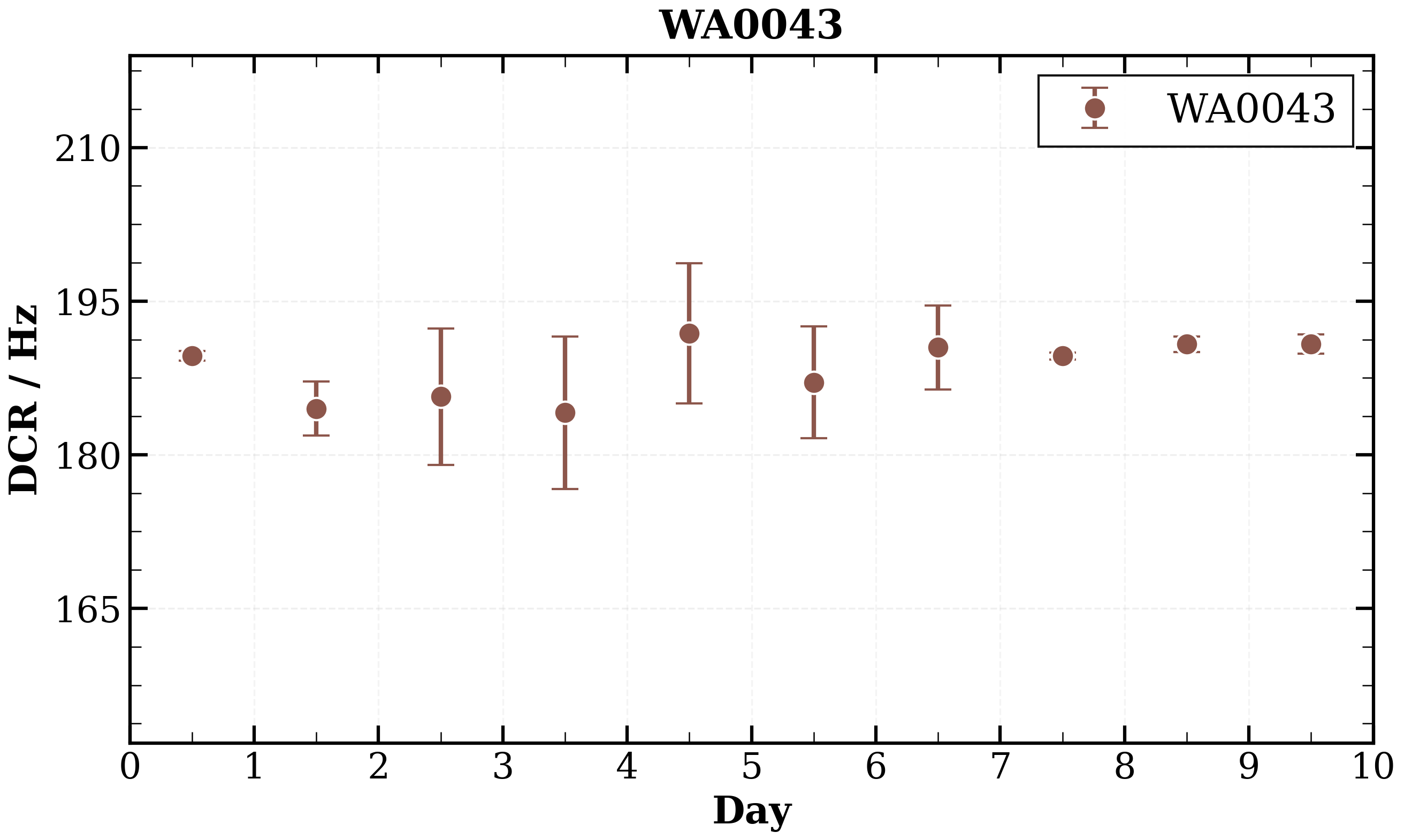}
    \end{minipage}
    \\[-0.7em]
    
    \begin{minipage}[b]{0.47\textwidth}
        \includegraphics[width=\textwidth]{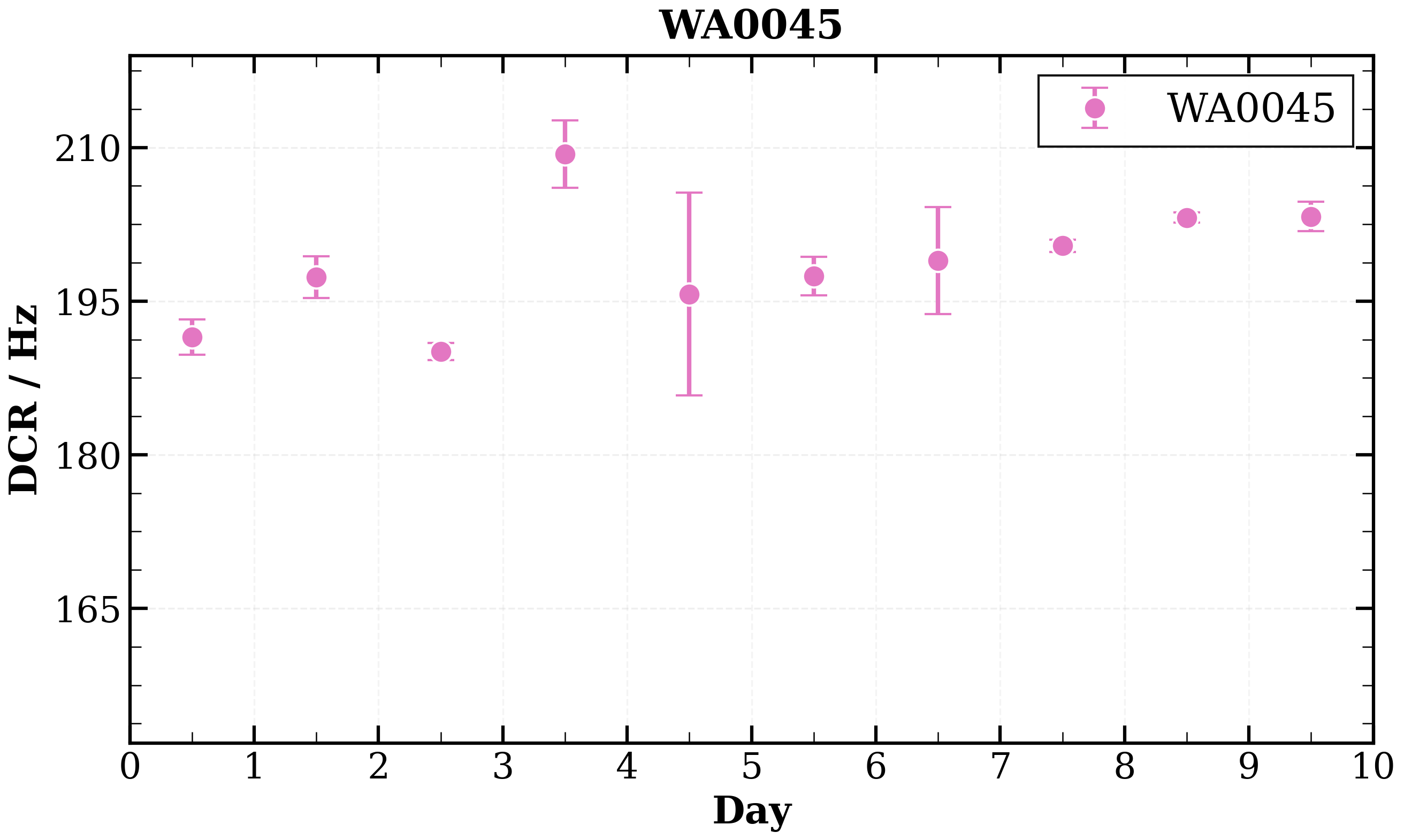}
    \end{minipage}
    
    \caption{Daily mean DCRs during the ten-day operating period in liquid argon at a motherboard input voltage of 900~V. Each point is plotted at the daily midpoint. The error bars represent the statistical standard deviation of repeated measurements within that day.}
    \label{fig:DCR_time_WA0037_45}
\end{figure}

The seven PMTs were operated continuously in liquid argon for ten days. Over the full ten-day period, the monitored system pressure had a mean of 15.000~psi and a standard deviation of 0.004~psi, indicating stable cryogenic operating conditions. Repeated measurements were averaged within each day to form the daily points shown in Figures~\ref{fig:Gain_Time_I900_WA0037_45} and \ref{fig:DCR_time_WA0037_45}. For the gain, each daily error bar is the quadrature sum defined in Equation~\ref{eq:gain_uncertainty}. For the DCR, each daily error bar is the statistical standard deviation of the repeated measurements.

\begin{table}[htbp]
  \centering
  \small
  \renewcommand{\arraystretch}{1.5}
  \begin{tabular}{lccccccc}
    \toprule
    PMT ID & WA0037 & WA0038 & WA0039 & WA0040 & WA0041 & WA0043 & WA0045 \\
    \midrule
    \makecell[l]{Gain / $10^{6}$}
    & \makecell[c]{$3.06$ \\ $\pm 0.03$ \\ $(0.92\%)$}
    & \makecell[c]{$3.41$ \\ $\pm 0.04$ \\ $(1.08\%)$}
    & \makecell[c]{$3.86$ \\ $\pm 0.05$ \\ $(1.19\%)$}
    & \makecell[c]{$2.47$ \\ $\pm 0.05$ \\ $(1.85\%)$}
    & \makecell[c]{$2.93$ \\ $\pm 0.03$ \\ $(0.98\%)$}
    & \makecell[c]{$2.71$ \\ $\pm 0.05$ \\ $(1.99\%)$}
    & \makecell[c]{$3.48$ \\ $\pm 0.07$ \\ $(2.04\%)$} \\
    \midrule
    \makecell[l]{DCR / Hz}
    & \makecell[c]{$187 \pm 2$ \\ $(1.13\%)$}
    & \makecell[c]{$180 \pm 6$ \\ $(3.29\%)$}
    & \makecell[c]{$189 \pm 3$ \\ $(1.71\%)$}
    & \makecell[c]{$166 \pm 4$ \\ $(2.20\%)$}
    & \makecell[c]{$188 \pm 4$ \\ $(2.14\%)$}
    & \makecell[c]{$188 \pm 3$ \\ $(1.52\%)$}
    & \makecell[c]{$199 \pm 6$ \\ $(2.88\%)$} \\
    \bottomrule
  \end{tabular}
  \captionsetup{width=\textwidth, justification=raggedright}
  \caption{Gain and DCR stability over the ten-day operating period. Each entry gives the mean and sample standard deviation of the ten daily means, calculated with a variance denominator of $n-1$, where $n=10$. The value in parentheses is the relative standard deviation (RSD), defined as the sample standard deviation divided by the mean and expressed as a percentage.}
  \label{tab:gain_dcr_stability}
\end{table}

Table~\ref{tab:gain_dcr_stability} summarizes the variation across the ten daily means using the sample standard deviation and RSD, distinct from the uncertainties shown for individual daily points. Across the seven PMTs, the gain RSDs range from 0.92\% to 2.04\%, while the DCR RSDs range from 1.13\% to 3.29\% over the ten-day observation period.

\section{Conclusion and outlook}\label{sec:4}

Seven Hamamatsu R8520-506 PMTs were characterized at room temperature and in liquid argon using the same motherboard and analysis procedures. At a motherboard input voltage of 900~V, the extracted gains in liquid argon are approximately 27.8\%--51.3\% higher than their room-temperature values. For all seven tubes, the P/V ratios increase and the relative SPE charge width, $\sigma_1/\mu_1$, decreases by approximately 5.3\%--14.4\%, indicating improved SPE charge resolution. No common direction is observed in the DCR changes across the sample: three DCR values increase and four decrease, with ranges of approximately 143--252~Hz at room temperature and 166--199~Hz in liquid argon. The afterpulse rates decrease by approximately 21.0\%--63.6\% for all seven PMTs. During ten days of continuous operation in liquid argon, the RSDs across the ten daily means of gain and DCR do not exceed approximately 2.1\% and 3.3\%, respectively.

The measured SPE response, afterpulse characteristics, and ten-day gain and DCR stability provide a basis for considering the R8520-506 as a photosensor candidate for future liquid-argon detectors, including DarkSide-LowMass~\cite{Agnes2023DarkSideLowMass}.

Further evaluation should increase the sample size, repeat the measurements across thermal cycles, and extend continuous operation beyond ten days. Tests with an integrated liquid-argon light-readout system would assess the suitability of these PMTs under detector-relevant optical and background conditions. A quantitative identification of the afterpulse ion species would additionally require information on the internal electric field and ion-production locations that is not available in the present analysis.

\backmatter

\bmhead{Acknowledgements}

This work is supported by the National Key Research and Development Project of China, Grant No.~2022YFA1602001.

\section*{Declarations}

\textbf{Conflict of interest} The authors declare no competing interests.

\bibliography{reference}

@article{Chepel2013,
  author        = {Chepel, Vitaly and Ara{\'u}jo, Henrique},
  title         = {Liquid noble gas detectors for low energy particle physics},
  journal       = {Journal of Instrumentation},
  volume        = {8},
  number        = {04},
  pages         = {R04001},
  year          = {2013},
  doi           = {10.1088/1748-0221/8/04/R04001},
  eprint        = {1207.2292},
  archivePrefix = {arXiv},
  primaryClass  = {physics.ins-det}
}

@article{AprileDoke2010,
  author        = {Aprile, Elena and Doke, T.},
  title         = {Liquid xenon detectors for particle physics and astrophysics},
  journal       = {Reviews of Modern Physics},
  volume        = {82},
  number        = {3},
  pages         = {2053--2097},
  year          = {2010},
  doi           = {10.1103/RevModPhys.82.2053},
  eprint        = {0910.4956},
  archivePrefix = {arXiv},
  primaryClass  = {physics.ins-det}
}

@techreport{HamamatsuR8520406001,
  author      = {{Hamamatsu Photonics K.K.}},
  title       = {Photomultiplier Tube {R8520-406-001}},
  institution = {Hamamatsu Photonics K.K.},
  type        = {Data sheet},
  number      = {TPMH1342E06},
  month       = sep,
  year        = {2024},
  url         = {https://www.hamamatsu.com/jp/en/product/optical-sensors/pmt/pmt_tube-alone/metal-package-type/R8520-406-001.html},
  note        = {Accessed 2 September 2026}
}

@article{Li2016PandaX,
  author        = {Li, Shaoli and Chen, Xun and Giboni, Karl L. and
                   Guo, Guodong and Ji, Xiangdong and Lin, Qing and
                   Liu, Jianglai and Mao, Yajun and Ni, Kaixuan and
                   Ren, Xiangxiang and Tan, Andi and Xiao, Mengjiao and
                   Xiao, Xiang and Zhou, Xiaopeng},
  title         = {Performance of photosensors in the {PandaX-I} experiment},
  journal       = {Journal of Instrumentation},
  volume        = {11},
  number        = {02},
  pages         = {T02005},
  year          = {2016},
  doi           = {10.1088/1748-0221/11/02/T02005},
  eprint        = {1511.06223},
  archivePrefix = {arXiv},
  primaryClass  = {physics.ins-det}
}

@article{Acciarri2012R11065,
  author        = {Acciarri, R. and Antonello, M. and Boffelli, F. and
                   Cambiaghi, M. and Canci, N. and Cavanna, F. and
                   Cocco, A. G. and Deniskina, N. and Di Pompeo, F. and
                   Fiorillo, G. and Galbiati, C. and Grandi, L. and
                   Kryczynski, P. and Meng, G. and Montanari, C. and
                   Palamara, O. and Pandola, L. and Perfetto, F. and
                   Piano Mortari, G. B. and Pietropaolo, F. and
                   Raselli, G. L. and Rubbia, C. and
                   Segreto, E. and Szelc, A. M. and Triossi, A. and
                   Ventura, S. and Vignoli, C. and Zani, A.},
  title         = {Demonstration and comparison of photomultiplier tubes at liquid {Argon} temperature},
  journal       = {Journal of Instrumentation},
  volume        = {7},
  number        = {01},
  pages         = {P01016},
  year          = {2012},
  doi           = {10.1088/1748-0221/7/01/P01016},
  eprint        = {1108.5584},
  archivePrefix = {arXiv},
  primaryClass  = {physics.ins-det}
}

@article{Canci2020DS50,
  author        = {Canci, N.},
  title         = {Long term operation with the {DarkSide-50} detector},
  journal       = {Journal of Instrumentation},
  volume        = {15},
  number        = {03},
  pages         = {C03026},
  year          = {2020},
  doi           = {10.1088/1748-0221/15/03/C03026},
  eprint        = {1912.05461},
  archivePrefix = {arXiv},
  primaryClass  = {astro-ph.IM},
  note          = {On behalf of the DarkSide-50 Collaboration}
}

@article{Baudis2013R11410,
  author        = {Baudis, Laura and Behrens, Annika and Ferella, Alfredo and
                   Kish, Alexander and Marrod{\'a}n Undagoitia, Teresa and
                   Mayani, Daniel and Schumann, Marc},
  title         = {Performance of the Hamamatsu {R11410} photomultiplier tube in cryogenic xenon environments},
  journal       = {Journal of Instrumentation},
  volume        = {8},
  number        = {04},
  pages         = {P04026},
  year          = {2013},
  doi           = {10.1088/1748-0221/8/04/P04026},
  eprint        = {1303.0226},
  archivePrefix = {arXiv},
  primaryClass  = {astro-ph.IM}
}

@article{E_Aprile_2012,
  author  = {Aprile, E. and Beck, M. and Bokeloh, K. and Budnik, R. and
             Choi, B. and Contreras, H. A. and Giboni, K.-L. and
             Goetzke, L. W. and Lang, R. F. and Lim, K. E. and
             Melgarejo Fernandez, A. J. and Plante, G. and Rizzo, A. and
             Shagin, P. and Weinheimer, C.},
  title   = {Measurement of the quantum efficiency of Hamamatsu {R8520} photomultipliers at liquid xenon temperature},
  journal = {Journal of Instrumentation},
  volume  = {7},
  number  = {10},
  pages   = {P10005},
  year    = {2012},
  doi     = {10.1088/1748-0221/7/10/P10005}
}

@article{Barrow_2017,
  author  = {Barrow, P. and Baudis, L. and Cichon, D. and Danisch, M. and
             Franco, D. and Kaether, F. and Kish, A. and Lindner, M. and
             Marrod{\'a}n Undagoitia, T. and Mayani, D. and Rauch, L. and
             Wei, Y. and Wulf, J.},
  title   = {Qualification tests of the {R11410-21} photomultiplier tubes for the {XENON1T} detector},
  journal = {Journal of Instrumentation},
  volume  = {12},
  number  = {01},
  pages   = {P01024},
  year    = {2017},
  doi     = {10.1088/1748-0221/12/01/P01024}
}

@article{adrover2025characterization,
  author  = {Adrover, M. and Baudis, L. and Bismark, A. and Colijn, A. P. and
             Cuenca-Garc{\'\i}a, J. J. and Decowski, M. P. and Flierman, M. and
             den Hollander, T.},
  title   = {Characterization of the Hamamatsu {R12699-406-M4} photomultiplier tube in cold xenon environments},
  journal = {Journal of Instrumentation},
  volume  = {20},
  number  = {12},
  pages   = {P12021},
  year    = {2025},
  doi     = {10.1088/1748-0221/20/12/P12021}
}

@misc{HamamatsuR8520506,
  author = {{Hamamatsu Photonics K.K.}},
  title  = {{Photomultiplier Tube R8520-406/R8520-506}}
}

@article{Agnes2023DarkSideLowMass,
  author        = {Agnes, P. and others},
  title         = {Sensitivity projections for a dual-phase argon {TPC} optimized for light dark matter searches through the ionization channel},
  journal       = {Physical Review D},
  volume        = {107},
  number        = {11},
  pages         = {112006},
  year          = {2023},
  doi           = {10.1103/PhysRevD.107.112006},
  eprint        = {2209.01177},
  archivePrefix = {arXiv},
  primaryClass  = {physics.ins-det}
}
\end{document}